\PassOptionsToPackage{table,xcdraw}{xcolor}
\documentclass[sigconf]{acmart}
\renewcommand\footnotetextcopyrightpermission[1]{}
\setcopyright{none}
\acmConference[]{}{}{}
\acmBooktitle{}
\acmPrice{}
\acmISBN{}
\acmDOI{}

\usepackage{tikz}
\usepackage{amsmath}
\usepackage{mdframed}
\usepackage[normalem]{ulem}
\usepackage{url}
\usepackage{makecell}
\usepackage{float}
\usepackage{enumitem} 

\usepackage{booktabs}
\usepackage{bbding}
\usepackage{colortbl}

\usepackage[most]{tcolorbox}
\newtcolorbox[auto counter, number within=section]{takeaway}[2][]{colframe=blue!80!black, colback=blue!10, coltitle=black, fonttitle=\bfseries, title=Takeaway~\thetcbcounter: #2,#1}

\iftrue
 \newcommand{\mannat}[1]{\textcolor{purple}{\textbf{[Comment: }\textit{#1} -- mannat\textbf{]}}}
    \newcommand{\ha}[1]{\textcolor{red}{\textbf{[Comment: }\textit{#1} -- ha\textbf{]}}}
     \newcommand{\sana}[1]{\textcolor{blue}{\textbf{[Comment: }\textit{#1} -- sana\textbf{]}}}

\else
      \newcommand{\sana}[1]{\ignorespaces}
      \newcommand{\ha}[1]{\ignorespaces}
       \newcommand{\mannat}[1]{\ignorespaces}
\fi

\renewenvironment{quote}
  {\list{}%
    {\setlength{\leftmargin}{10pt}
     \setlength{\rightmargin}{0pt}
     \setlength{\topsep}{2pt}
     \setlength{\partopsep}{0pt}%
     \setlength{\parskip}{0pt}%
     \setlength{\parsep}{2pt}}%
   \item\relax}
  {\endlist}

\begin{document}


\title[Exploring Privacy Perspectives of Indian Internet Users in Light of DPDPA]{``Nobody should control the end user'': Exploring Privacy Perspectives of Indian Internet Users in Light of DPDPA}


\author{Sana Athar}
\affiliation{%
  \institution{MPI-INF}
  \city{}
  \country{}}
\email{sathar@mpi-inf.mpg.de}

\author{Devashish Gosain}
\affiliation{%
  \institution{IIT Bombay}
  \city{}
  \country{}}
\email{dgosain@cse.iitb.ac.in}

\author{Anja Feldmann}
\affiliation{%
  \institution{MPI-INF}
  \city{}
  \country{}}
\email{anja@mpi-inf.mpg.de}

\author{Mannat Kaur}
\affiliation{%
  \institution{MPI-INF}
  \city{}
  \country{}}
\email{mkaur@mpi-inf.mpg.de}

\author{Ha Dao}
\affiliation{%
  \institution{MPI-INF}
  \city{}
  \country{}}
\email{hadao@mpi-inf.mpg.de}







\renewcommand{\shortauthors}{Athar et al.}

\begin{abstract}
With the rapid increase in online interactions, concerns over data privacy and transparency of data processing practices have become more pronounced.
While regulations like the GDPR have driven the widespread adoption of cookie banners in the EU, India’s Digital Personal Data Protection Act (DPDPA) promises similar changes domestically, aiming to introduce a framework for data protection. 
However, certain clauses within the DPDPA raise concerns about potential infringements on user privacy, given the exemptions for government accountability and user consent requirements. 
In this study, for the first time, we explore Indian Internet users' awareness and perceptions
of cookie banners, online privacy, and privacy regulations, especially in light of the newly passed DPDPA. 
We conducted an online anonymous survey with 428 Indian participants,
which addressed:
\textbf{(1)} users' \emph{perspectives} on
cookie banners, \textbf{(2)} their 
attitudes towards online privacy and privacy regulations, and \textbf{(3)} their acceptance of 10 contentious DPDPA clauses that favor state authorities and may enable surveillance.
Our findings reveal that privacy-conscious users often lack consistent awareness of privacy mechanisms, and their concerns do not always lead to protective actions. 
Our thematic analysis of 143 open-ended responses shows that users' privacy and data protection concerns are rooted in skepticism towards the government, shaping their perceptions of the DPDPA and fueling demands for policy revisions.
Our study highlights the need for clearer communication regarding the DPDPA, user-centric consent mechanisms, and policy refinements to enhance data privacy practices in India.

\end{abstract}




\maketitle

\section{Introduction} 
 
The Internet has become a vital platform for communication, e-commerce, and banking, resulting in the collection of vast amounts of personal data by websites and online services. Users, often unaware of how much personal data they share, become the subject of online tracking and user profiling \cite{acquisti2006imagined, yu2016tracking, agarwal2013not, krasnova2009privacy}.
Thus, several countries worldwide have introduced regulations to protect user privacy in response to these concerns.
Global frameworks, such as the General Data Protection Regulation (GDPR) \cite{GDPR} in the EU and the California Consumer Privacy Act (CCPA) \cite{CCPA} in the U.S., have established benchmarks for data protection laws worldwide. 
One visible outcome of these regulations is the prevalence of cookie consent banners, which inform users about data tracking and request their consent to proceed. While these banners are prominent in the EU, they are relatively less prevalent in countries like India \cite{rasaii2023exploring}, where comprehensive privacy laws are still in the process of emerging.

India's expanding digital ecosystem and large Internet user base \cite{indiagrowthreport2024} drive growth in digital transactions, social media, and e-commerce. Yet, many users remain unaware of the risks associated with data collection and the protections in place for their privacy \cite{norton, PwC2024DPDPAsurvey}.
Recognizing this gap, the approval of the Digital Personal Data Protection Act (DPDPA) by the Indian Parliament in 2023 marks a significant step towards stronger data privacy laws in India \cite{dpdpa2023}.
Similar to the GDPR, the DPDPA emphasizes user consent and regulates the collection, storage, and processing of personal data. It mandates businesses to obtain explicit consent and transparency but raises concerns over government exemptions and vague terms that may threaten privacy \cite{TheHindu, sundara2023digital} (see \autoref{group4}). These ambiguities create uncertainty in implementation and potential misuse. Thus, understanding user sentiment is crucial for assessing its effectiveness, fostering trust, and ensuring it meets expectations for strong data protection.

Furthermore, Indian users remain underrepresented in privacy research, even though their perspectives -- distinct from European contexts -- are crucial as the DPDPA ushers in a new era of digital privacy.
Broad and diverse participant samples in studies like the survey by PwC-India \cite{PwC2024DPDPAsurvey} show awareness of DPDPA is currently concentrated among educated and digitally literate urban groups—with limited exposure among the wider population.
Once the DPDPA is enforced, cookie banners and similar mechanisms are expected to become more prevalent on websites popular in India. Thus, understanding how Indian users engage with these banners and their attitudes toward privacy-infringing clauses of the DPDPA is crucial. 

To address these gaps, we conducted an online survey with 428 participants, who are likely familiar with online environments due to their educational background.
Our analysis focuses on reported interactions with cookie banners, perceptions of online privacy, and acceptance of DPDPA clauses. 
The survey design included questions divided into four groups---demographics, cookie banner experiences, online privacy \emph{perspectives}, and acceptance of DPDPA clauses---and an open-ended feedback section, which was analyzed using thematic analysis (\autoref{sec:method}).
The study offers insights into how Indians engage with these privacy-related features online and addresses the following research questions. 

\vspace{2mm}
\begin{mdframed}
\textbf{RQ1}: 
What are the \emph{perspectives} of Indian users on encountering and managing cookie banners?
\end{mdframed} 
\vspace{2mm}

Our analysis reveals that 91\% of respondents reported encountering a cookie banner while browsing the web, yet nearly 50\% reported difficulty locating and interacting with the ``Reject'' option.
In addition, the key factors influencing their interactions include banner category (the design and options presented in the banner), disclaimer notice (the clarity and visibility of the banner’s message), website context (the type of the website), and awareness levels (limited understanding of data privacy implications) (see \autoref{subsec:rq1}). 

\vspace{2mm}
\begin{mdframed}
\textbf{RQ2}: 
To what extent do Indian users value browsing privacy, and how do their expressed concerns align with their reported consent actions?
\end{mdframed} 
\vspace{2mm}

We find that 82.2\% of respondents identify as privacy-conscious regarding their browsing history, yet a significant gap exists in online privacy knowledge. 
While users express concerns about privacy and data handling, reflecting awareness, this concern often fails to translate into informed actions, such as rejecting cookies, revealing an evident privacy paradox.
The key factors for this include unawareness (limited understanding of consent mechanisms and privacy regulations) and misconceptions (incorrect assumptions about how personal data is used or protected) (see \autoref{subsec:rq2}).

\vspace{2mm}
\begin{mdframed}
\textbf{RQ3}: What is the acceptance level of Indian users towards different clauses of DPDPA?
\end{mdframed} 
\vspace{2mm}

Our findings reveal a mixed level of acceptance among Indian users of the DPDPA clauses, which seem to favor state authorities and potentially enable surveillance.
Clauses related to national security received relatively higher acceptance, which was explained by our thematic analysis. Users demand no exemptions for anyone in the law unless it's a national emergency. In contrast, a clause dissuading lawsuits against the Government was deemed least acceptable. This was supported by qualitative responses highlighting a demand for accountability, with participants asserting no authority should be exempt from responsibility under the law (see~\autoref{subsec:rq3}).

Notably, our \textit{thematic analysis} \cite{clarke2017thematic} uncovered three themes: respondents' significant concerns about online privacy, data protection, and cookies (T1), emphasizing the need for structural improvements in the DPDPA framework (T2) while expressing distrust in the government’s intentions and ability to handle data responsibly (T3). The sentiment of \textbf{distrust} towards the Government is a critical concept since it fuels users' concerns about personal data privacy and determines how they perceive DPDPA. These themes are interrelated, with privacy concerns driving demands for revisions in DPDPA, both shaped by skepticism about government motives and capabilities (see~\autoref{subsec:openfield}). Finally, we discuss our limitations and compare our key insights with related work in \autoref{sec:discussion}.
\section{Background and Related Work} 

This section summarizes prior works along three domains related to our research questions: online privacy and regulations, cookie banner interactions, and user privacy perceptions.  

\subsection{Online privacy and regulations}
Online tracking, which involves collecting users' online behavior data, raises concerns about privacy and data use~\cite{pavlou2011state, lee2011personalisation, belanger2002trustworthiness}. 
To address these concerns, regulations like the GDPR \cite{GDPR} and CCPA \cite{CCPA} govern web cookies and tracking technologies. While they enhance data protection, user awareness and perceptions vary. Degeling et al. \cite{degeling2018we} found that GDPR improved transparency and consent adoption but noted compliance gaps and widespread ``consent fatigue'' from cookie banners.
Zaeem et al. \cite{zaeem2020effect} and Gáti et al. \cite{gati2020perception} highlighted innovative privacy policy designs to meet GDPR’s transparency and informed consent goals. Mangini et al. \cite{mangini2020empirical} found users often know their Right to be Forgotten (RTBF) but remain confused about its scope, while resource-limited organizations struggle with GDPR compliance, including RTBF requests.
Kyi et al. \cite{kyi2024doesn} found a gap between user expectations and privacy disclosures, where vague legal language causes confusion and mistrust. Strycharz et al. \cite{strycharz2020data} noted that despite GDPR, many users still fear data misuse. Solove \cite{solove2012introduction} critiques consent-based protection, highlighting issues like complex policies, user fatigue, and power imbalances.

While substantial research focused on different aspects of existing privacy laws \cite{o2021clear, chen2021fighting, mangini2020empirical, kyi2024doesn}, DPDPA \cite{dpdpa2023} is relatively new, yet to be enacted and not explored. The DPDPA was passed as a law in India in 2023, establishing guidelines for the processing and protection of digital personal data of Indian citizens, with a strong emphasis on user consent. Although the DPDPA aims to enhance data protection, its provisions contain ambiguous terms that could create uncertainty in interpreting key clauses~\cite{sundara2023digital}. These ambiguities, particularly in the exceptions granted to government agencies, raise concerns about potential loopholes that might be exploited to justify mass surveillance or weaken user privacy safeguards.

\begin{figure*}[t!]
    \centering    
    \includegraphics[width=\linewidth]{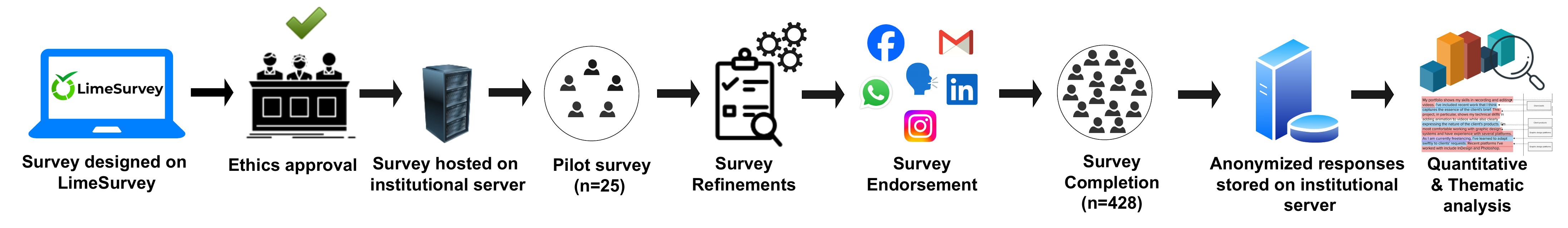}
    \vspace{-3mm}
    \caption{Overview of our methodology.} 
    \label{fig:Overview}
\end{figure*}

\subsection{Cookie banner and user interactions}

Growing privacy concerns and data protection laws have led to the widespread use of cookie banners \cite{rasaii2023exploring, jha2022internet}. These banners inform users and request consent, with studies showing that design, wording, defaults, and visuals impact user engagement.
Boerman et al.~\cite{doi:10.1080/00913367.2017.1339368} found that complex language and lengthy cookie notices lead users to ignore or accept them without reading. Nouwens et al. \cite{nouwens2020dark} highlighted how ``dark patterns'' increase acceptance but undermine autonomy. Berens et al. \cite{berens2022cookie} showed that visual design, button labeling, and explanations influence user behavior.
Ma and Birrell \cite{ma2022prospective} found that negatively framed labels influence user decisions, aligning with prospect theory. Kulyk et al. \cite{kulyk2018website} noted a preference for bottom-page banners to minimize disruption, with options like \emph{Accept All} vs. \emph{Manage Preferences} shaping behavior. Mathur et al. \cite{mathur2019dark} highlighted that users prefer banners allowing easy cookie rejection.
Singh et al. \cite{singh2022cookie} found that users prefer simple, concise consent notices, favoring designs like sliders or categorized cookies.
Furthermore, user trust has a significant impact on cookie banner interactions. 
Schermer et al. \cite{schermer2014crisis} and Kulyk et al. \cite{kulyk2018website} found that users engage more with banners on trusted websites, influenced by data sensitivity and the services offered.
Chanchary and Chiasson \cite{chanchary2015user} found that privacy fatigue influences user behavior, leading to desensitization and automatic cookie acceptance. Utz et al. \cite{utz2019informed} observed that users engage more thoughtfully when given meaningful choices. Kulyk et al. \cite{kulyk2020has} noted that cookies are seen as intrusive, raising privacy concerns, providing insufficient information, and contributing to fatigue. 


\subsection{User privacy perceptions}
Online privacy remains a major concern for Internet users, yet a gap persists between their attitudes and behaviors.
The ``privacy paradox,'' a term coined by Acquisti et al. \cite{acquisti2016economics}, highlights the gap between users' stated privacy concerns and their contradictory actions, such as readily accepting cookies or sharing personal information. Barth and de Jong \cite{barth2017privacy} noted that users trade data for convenience, such as personalized content. Norberg et al. \cite{norberg2007privacy} found users often underestimate data-sharing risks, assuming companies will protect their information.
Brandimarte et al. \cite{brandimarte2013misplaced} found that perceived online anonymity leads to greater personal disclosure than face-to-face.
Stutzman et al. \cite{stutzman2013silent} observed privacy fatigue on Facebook, with users focusing on peer visibility while neglecting institutional and third-party risks.
Kokolakis \cite{kokolakis2017privacy} and Gerber et al. \cite{gerber2018explaining} identified privacy intention, attitude, willingness to disclose, and concern as key predictors of privacy behavior, with demography playing a minor role.
Hoffmann et al. \cite{hoffmann2016privacy} linked the privacy paradox to ``privacy cynicism,'' where distrust fosters disengagement, highlighting a gap between knowledge and behavior. P{\"o}tzsch \cite{potzsch2009privacy} argued that awareness alone is insufficient, stressing the need for intuitive privacy tools. Fleming et al. \cite{fleming2021role} highlighted factors like data value, trust, and cultural differences in shaping privacy behaviors, with cross-country variations limiting universal strategies \cite{sambasivan2018privacy}.


\subsection{Comparison with the related work} 
Our study differs from and extends the related work in the following ways: 
(1) We study how Indian users choose to interact with different types of cookie banners, by presenting them screenshots of actual banners with the full context (buttons, disclaimer, website, placement, dismiss option). Unlike previous studies on European/North American populations~\cite{kulyk2020has,singh2022cookie,gerber2018explaining,hoffmann2016privacy,mangini2020empirical,strycharz2020data,kyi2024doesn,gati2020perception}, we exclusively focus on an understudied population, likely with different tendencies and choices.
(2) We evaluate the general perception of Indian users towards online privacy vis-à-vis browsing history, cookie banners, and third-party tracking, all in one survey. 
By analyzing concerns and consent actions, we identify gaps reflecting the privacy paradox \cite{gerber2018explaining, hoffmann2016privacy}. This evaluation helps determine whether current privacy mechanisms meet user needs or require improvements.
(3) Lastly, our research is the \textit{first} academic study to examine Indian users' perceptions and acceptance of DPDPA, filling a crucial gap. To the best of our knowledge, no prior studies exist on this regulation and existing literature primarily focuses on user awareness and expectations rather than \emph{acceptance} of major privacy laws like GDPR or CCPA ~\cite{kulyk2020has,singh2022cookie,mangini2020empirical, strycharz2020data, gati2020perception}.


\section{Methodology} 
\label{sec:method}


This section outlines the survey design, implementation, and data analysis approach. The overview of our methodology is presented in ~\autoref{fig:Overview}.

\subsection{Ethical consideration} 
Ethics reflections are central to our study design. 
We applied for an ethics review of our survey study---including the study overview, the consent notice, and the plans for data collection, storage, processing, and deletion---to our University's ethics committee, for which we received approval.
We introduced the study purpose at the beginning of the survey, including the rights of the data subject and that participation was voluntary (see privacy notice ~\autoref{privacynotice}).
We conducted an anonymous survey using an internally-hosted instance of LimeSurvey \cite{Lime}. This ensured that we did not collect any personally identifiable information (PII) in our survey, and we also ensured that any PII (e.g., voluntarily provided email addresses) was removed before conducting our analysis. 
The collected data was stored on the institution's internal servers in compliance with its data security policies.

\subsection{Survey design} \label{subsec:surveydesign}
The survey included 38 questions, took about 15 minutes, and began with a research description, author affiliations, and a privacy notice (see \autoref{privacynotice}).

The privacy notice contained general information about (1) the demographic information we intended to collect, (2) the survey being anonymous as we did not seek PII, and (3) the participation being completely voluntary as respondents can opt-out anytime before clicking the ``submit'' button at the end. We then asked the respondents to confirm that they were over 18, had read and understood the privacy notice, and had volunteered to participate in the survey by checking a mandatory box. 
After consent, they answered survey questions, divided into four groups, with an open question at the end (see \autoref{group3}).

\vspace{1mm}

\noindent \textbf{Group 1}: \textit{About You}  \textit{(participant demographics)} - This group contained nine questions about basic demographic information of respondents like age group, gender, state, professional status, and academic background (which we report in an aggregated manner). We collected this information to get insights about the survey respondents. We did not ask for PII such as name, email, etc., making the survey anonymous. In addition, we added questions about Internet usage and online data-sharing preferences. 
\vspace{1mm}

\noindent \textbf{Group 2}: \textit{Your Experience} \textit{(participant reported experience with cookie banners)} - In this group, we had ten questions to infer if respondents had previously encountered cookie banners while browsing the Internet. 
Similar to \cite{kulyk2018website}, we presented the respondents with different types of banners as screenshots. We asked them to state their preferred way of interacting with banners (e.g., closing, accepting, rejecting them) when encountered while browsing the Internet. We also asked respondents how often they click on the ``Accept'' or ``Reject'' buttons on cookie banners whenever they find them. 
We aimed to understand respondents' perceptions of their interactions with cookie banners, specifically the choices they believe they would make when encountering different types of banners.
\vspace{1mm}

\noindent \textbf{Group 3}: \textit{Your Perspective} (\textit{participant perspectives about online privacy}) - 
This group included eight statements assessing participants' opinions on online data privacy. The statements explored ordinary Indians' awareness, attitudes, and views on data safety and the need for privacy regulations in India. Respondents rated their agreement on a 4-point Likert scale: Strongly Agree, Agree, Disagree, and Strongly Disagree.       
\vspace{1mm}

\noindent \textbf{Group 4}: \textit{Your Acceptance}  (\textit{participant reported acceptance towards DPDPA clauses}) - In this group, we presented the respondents with 10 clauses, selected from a total of 44 clauses from DPDPA, that seem to favor state authorities and may lead to surveillance. 
The decision to include these specific clauses in the study was informed by a broader perspective drawn from secondary sources, such as legal blogs \cite{DPDPAblogpost}, following the approval of the DPDPA.
Some legal experts highlighted issues related to privacy, transparency, and governmental accountability, pointing to specific clauses that seemed problematic \cite{sundara2023digital}. Some terms are unclear and left undefined in the documentation, which could lead to loopholes and exploitative implementation of the law. Thus, we carefully reviewed the official documentation of DPDPA and selected 10 clauses that contained exceptions that could potentially weaken the rights of citizens, empowering the authorities and business organizations to process personal data with minimal transparency and lacking adequate provisions for user protection and recourse. 
We asked respondents to state their level of acceptance towards them. 
We used a 4-point Likert scale with the following choices: Highly Acceptable, Acceptable, Unacceptable, Highly Unacceptable.
\vspace{1mm}

\noindent \textbf{\textit{Open-ended field for comments/feedback}}: At the end, we added an open-ended field for respondents to share their opinions on our research project, methodology, or DPDPA, which was not mandatory. We received 143 comments on this open-ended field, with at least 63 remarks appreciating our research initiative and indicating that this survey was timely, relevant, and informative for users regarding online privacy and DPDPA. We collected qualitative data through this open-ended field where people voluntarily shared their views, concerns, and preferences regarding cookie banners, online privacy, and DPDPA (see ~\autoref{group5}). 
Later, in ~\autoref{subsec:analysis}, we elaborate on our thematic analysis of the collected data.

\subsection{Survey implementation}
\textit{\textbf{Platform:}} 
We hosted an instance of LimeSurvey \cite{Lime} and stored the responses on our institution's internal servers. 
We chose LimeSurvey instead of other paid crowd-sourcing platforms, e.g., Amazon MTurk \cite{MTurk}, for several reasons. First, such US-based platforms are known to have privacy concerns~\cite{xia2017our},
and Prolific \cite{prolific}, a European alternative, did not have a large enough number of Indian participants at the time of our survey.
Hosting the data internally also allowed us to uphold the highest standards of scientific integrity and data protection by avoiding third-party involvement.
Second, we ensured that we did not contribute to known exploitative practices on crowd-sourcing platforms~\cite{virtualsweatshop}.
Third, a financially incentivized survey is prone to incentive bias and dishonest/random responses, which can compromise the data quality.
Thus, we used voluntary participation, which mitigated these risks.
Lastly, participants recruited via mainstream paid platforms raise concerns about non-naiveté \cite{chandler2014nonnaivete}, as their familiarity with surveys and typical research methods may skew their responses.
Hence, our approach engaged the general population, yielding more candid and naive responses. Limitations of this approach are discussed in 
\autoref{sec:limitations}.
\vspace{1mm}

\noindent \textit{\textbf{Pilot test:}} 
To enhance the user experience, quality, and readability of the survey, we conducted a pilot test with 25 participants recruited via LinkedIn. These participants expressed interest in the survey and provided genuine feedback. Based on their input, we implemented two major modifications.
\vspace{1mm}

\noindent \textit{\textbf{Refinements:}} 
The first key feedback highlighted compatibility issues in the question formats of Groups 3 and 4 (\autoref{group4}), where the survey's user interface did not render properly on mobile devices, limiting participation. To address this, we modified the answer format and introduced a four-category scale.
The second major feedback concerned the complexity of DPDPA clauses in Group 4, where legal terminology made comprehension difficult. We paraphrased these clauses for clarity while referencing the original chapter and clause numbers for those seeking detailed documentation.
The rest of the survey remained unchanged, as no other significant issues were reported.

\vspace{1mm}

\noindent \textit{\textbf{Recruitment strategy:}}  We designed a flyer (see flyer ~\autoref{flyer}) containing our project description and survey link. To collect responses, we posted and forwarded the link and flyer on social media platforms (LinkedIn, Facebook, WhatsApp, and Instagram). We sent 135 cold emails to people working in different Universities all over India. Lastly, we shared the survey across our networks and requested that they 
forward it to their social groups. We discuss the limitations of our approach and the mitigation strategies in \S~\ref{sec:limitations}.

\subsection{Data analysis}  \label{subsec:analysis}
We received 428 complete responses over 40 days (May–June 2024). Anonymized data was exported from LimeSurvey and analyzed individually and cross-referenced to address the three research questions.


To answer RQ1, we analyzed the respondents' preferred way of interacting with the various cookie banners that we presented them in Group 2 (\autoref{group2}) of our survey (see \S~\ref{subsec:rq1}).

For RQ2, we evaluate respondents' perception of online privacy based on their (dis)agreement with statements presented to them in Group 3 (\autoref{group3}). We further compare respondents' claims to choices between questions from Groups 1, 2, and 3 to identify any gaps or contradictions. This enabled us to match respondents' privacy perceptions with their eventual choices regarding data privacy in survey questions (see \autoref{subsec:rq2}).

For RQ3, we performed a Kruskal Wallis \cite{kruskal1952} test to identify any significant differences in the acceptance levels of all clauses. For this, we considered the null hypothesis, i.e., \textit{H0:} There are no significant differences in the acceptance level among the ten clauses, with $\alpha = 0.01$. 
To perform this test, we mapped the Likert scale responses (in Group 4) to different numerical ranks: Highly Unacceptable (1), Unacceptable (2), Acceptable (3), and Highly Acceptable (4). 
We also determined mode values for the acceptance level of every clause using these ranks to find the most preferred reported acceptance level of respondents towards each clause (see~\autoref{subsec:rq3}).

Lastly, we used the \textbf{thematic analysis} (TA) \cite{clarke2017thematic} approach to analyze the qualitative responses (143 entries, see \autoref{group5}) obtained through the open-ended field. 
We carried out qualitative coding using the free and open-source tool \textit{Taguette} \cite{Taguette}. 
We carefully checked all the qualitative responses for PII and de-identified them before uploading them to open-source platforms.\footnote{One participant voluntarily added their email ID, which was removed.}
Two researchers collaboratively assigned codes to the responses over four iterations, resulting in 48 codes (see codebook ~\autoref{codebook}), finalized after discussion with the research team. 
We then systematically clustered these codes into broader groups based on conceptual similarities using \textit{Miro}, a digital collaborative platform \cite{Miro}.
After refining these groups, we generated three distinct themes that reflected the core concerns and insights shared by respondents. 
Although each theme reflects distinct ideas, they are closely interrelated, forming a comprehensive narrative about how respondents view online privacy and the role of Government and privacy regulations in safeguarding personal data. 
The themes are enumerated as follows:

\begin{enumerate}
    \item[T1:] \textit{User Concerns about Privacy, Data Protection, and Cookies:} This theme captures respondents' insights regarding current privacy risks, concerns regarding protecting their personal data online, and skepticism about the use of cookies on websites.
    \item[T2:] \textit{Improving the DPDPA Framework:} This theme encompasses the general pessimism and constructive criticism surrounding the current state of DPDPA, and suggestions that could improve the framework.
    \item[T3:] \textit{Trust-Distrust in the System:} This theme highlights the various levels of trust that respondents have expressed towards the Judiciary and Government, including instances of distrust and skepticism towards the Government. 
\end{enumerate}
In line with the reflexive TA, we reflected on our role as researchers. Our team consisted of three Indian researchers hailing from different Indian states. In addition, our team included a Vietnamese and a German researcher. Two of the researchers are experts in online privacy and privacy regulations, one is an expert in human factors research, and one is an expert in Internet infrastructure. Our diverse and multidisciplinary team was well-suited to conduct this research and interpret the findings within the necessary local, global, and socio-technical contexts.


\section{Results}

\noindent \textbf{Participant profile:}
Out of 428 complete responses received, almost 65\% were from men and 35\% from women. 
The most common age group was 18--25 years (38.5\%), followed by 26--35 (25.7\%), with the remaining respondents belonging to older age groups. Students comprised 37.6\% of the respondents, followed by industry professionals at 29.6\%, academic researchers at 17\%, and others at 15.6\%. 
A notable proportion of the respondents had backgrounds in engineering (36.6\%), followed by business (26.8\%), computer sciences (12.1\%), theoretical sciences (6.3\%), social sciences (6.3\%), medical sciences (4.4\%), and other fields (7.2\%) (see~\autoref{tab:demographics}).

\begin{table}[h!]
    \centering
    \rowcolors{2}{gray!10}{white}
    \resizebox{\linewidth}{!}{%
    \begin{tabular}{lrr}
        \toprule
        \textbf{Gender} & \textbf{\#} & \textbf{\%} \\
        \midrule
        Men & 278 & 64.9 \\
        Women & 150 & 35.0 \\
        \midrule
        \multicolumn{3}{l}{\textbf{Age Groups}} \\
        \midrule
        18-25 & 165 & 38.5 \\
        26-35 & 110 & 25.7 \\
        36-45 & 42 & 9.8 \\
        46-55 & 71 & 16.5 \\
        56-65 & 37 & 8.6 \\
        66-75 & 3 & 0.7 \\
        Above 75 & 0 & 0.0 \\
        \midrule
        \multicolumn{3}{l}{\textbf{Profession}} \\
        \midrule
        Student & 161 & 37.6 \\
        Academic Researcher & 73 & 17.0 \\
        Industry Professional & 127 & 29.6 \\
        Other & 67 & 15.6 \\
        \midrule
        \rowcolor{white}\multicolumn{3}{l}{\textbf{Field of Work/Research}} \\
        \midrule
        Computer Science & 52 & 12.1 \\
        Mathematics/Statistics/Physics/Chemistry or other Science & 27 & 6.3 \\
        Engineering and related fields & 157 & 36.6 \\
        Medical Sciences and related fields & 19 & 4.4 \\
        Humanities/Law/Journalism and related fields & 27 & 6.3 \\
        Commerce/Finance/Economics/Business/Management or related fields & 115 & 26.8 \\
        Other & 31 & 7.2 \\
        \bottomrule
    \end{tabular}
    }
    \caption{Participant demographics (n=428).}
    \label{tab:demographics}
\end{table}

\begin{table*}[t!]
\centering
\resizebox{\linewidth}{!}{%
\begin{tabular}{|llllll|lll|lll|}
\hline
\multicolumn{3}{|c|}{\textbf{B1.1: Confirmation only}}                                               & \multicolumn{3}{c|}{\textbf{B1.2: Confirmation only}}                                              & \multicolumn{3}{c|}{\textbf{B2: Binary}}                                       & \multicolumn{3}{c|}{\textbf{B3: No option}}                                                  \\ \hline
\multicolumn{1}{|l|}{``Agree and Close"}    & \multicolumn{1}{l|}{198} & \multicolumn{1}{l|}{46.2\%} & \multicolumn{1}{l|}{``Accept Cookies"}                         & \multicolumn{1}{l|}{130} & 30.3\% & \multicolumn{1}{l|}{``Consent"}            & \multicolumn{1}{l|}{46}  & 10.7\% & \multicolumn{1}{l|}{``Got it"}                           & \multicolumn{1}{l|}{95}  & 22.1\%  \\ \hline
\multicolumn{1}{|l|}{``Learn More"}         & \multicolumn{1}{l|}{101} & \multicolumn{1}{l|}{23.5\%} & \multicolumn{1}{l|}{``I ignore the banner continue browsing''} & \multicolumn{1}{l|}{120} & 28.0\%   & \multicolumn{1}{l|}{``Do not consent"}     & \multicolumn{1}{l|}{191} & 44.6\% & \multicolumn{1}{l|}{``I exit the website"}               & \multicolumn{1}{l|}{134} & 31.3\% \\ \hline
\multicolumn{1}{|l|}{``I exit the website"} & \multicolumn{1}{l|}{129} & \multicolumn{1}{l|}{30.1\%} & \multicolumn{1}{l|}{``I close the banner''}                    & \multicolumn{1}{l|}{178} & 41.5\% & \multicolumn{1}{l|}{``Manage options"}     & \multicolumn{1}{l|}{47}  & 10.9\% & \multicolumn{1}{l|}{``I avoid visiting such a website''} & \multicolumn{1}{l|}{199} & 46.4\% \\ \hline
\multicolumn{6}{|l|}{}                                                                                                                                                                                    & \multicolumn{1}{l|}{``I close the banner"} & \multicolumn{1}{l|}{144} & 33.6\% & \multicolumn{3}{l|}{}                                                                        \\ \hline
\end{tabular}
}
\caption{Users' choice of interaction with different types of cookie banners (Confirmation only, Binary and No option).
    }
    \label{tab:banners}
\end{table*}

Nearly 52\% (225/428) of the responses came from one state in India, Uttar Pradesh (UP) (see \autoref{tab:states} in the~\autoref{subsec:states}). This can be attributed to our recruitment strategy, which leveraged the personal networks of the authors. To check whether the geographical skew disproportionately affected the responses, we isolated the UP responses from the rest of the data. We observed that the number of respondents choosing each option in every question was nearly halved. For example (see G03-Q.4
\footnote{We annotate the survey questions by combining the group number with the question number: G0i-Q.j, where i = group number and j = question number.} 
in~\autoref{questionnaire}): ``I am fine with websites using third-party cookies for tracking/profiling me across the Internet''. The responses, including all participants, were: (i) Strongly agree (150), (ii) Agree (55), (iii) Disagree (206), and (iv) Strongly disagree (17). The responses, excluding the UP respondents, were: Strongly Agree (80), Agree (31), Disagree (105), and Strongly Disagree (9).
This indicates that the responses of the UP participants did not significantly alter the trends visible in our findings.
We observed a similar effect for all other questions (see~\autoref{sec:limitations} for details).

\subsection{RQ1: What are the perspectives of Indian users on encountering and managing cookie banners?}
\label{subsec:rq1}

Indian users are primarily exposed to cookie banners on international websites\footnote{These websites implement banners likely due to privacy laws enforced in other regions, such as GDPR in the EU and CCPA in California \cite{rasaii2023exploring}.}.
However, in India, the DPDPA law has been approved by parliament but has not yet been enacted into law. Once DPDPA is enforced, cookie banners are expected to become more common across Indian websites. Understanding how Indian users perceive their interaction with these banners will help assess their readiness for increased exposure to these consent notices. To accomplish this, we presented the respondents with screenshots of different types of cookie banners and asked them to choose their interaction with them (see Group 2 questions in our survey~\autoref{group2} for details).

\begin{figure}[h!]
    \centering
    \includegraphics[width=0.45\textwidth]{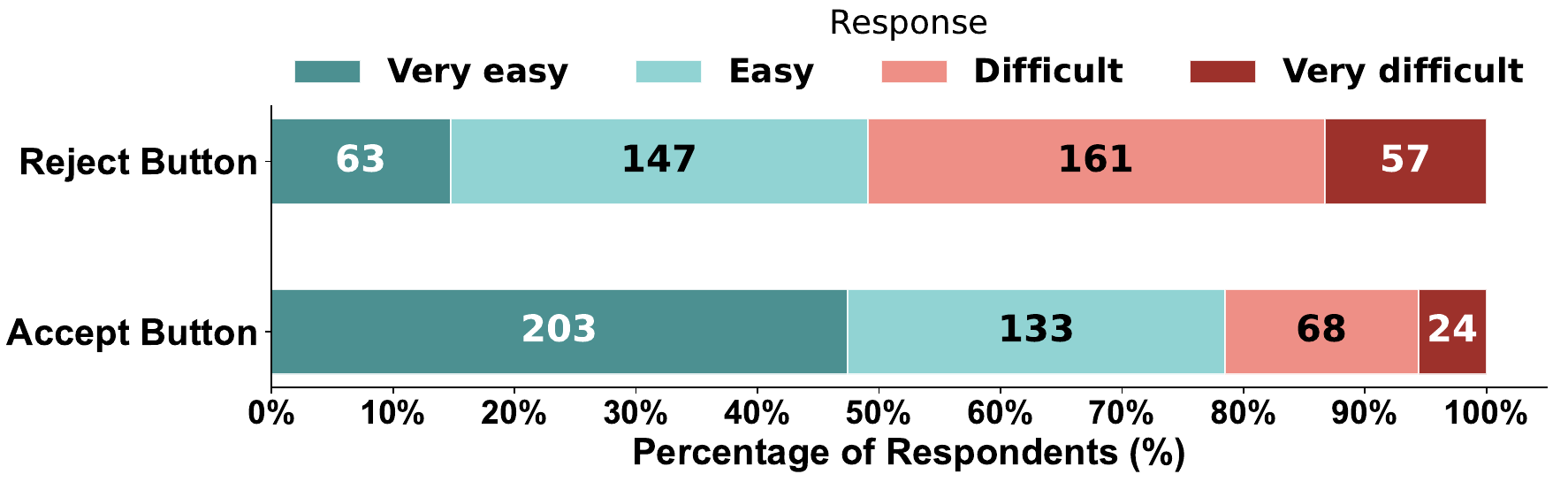}
    \caption{Ease of finding and clicking on \emph{Reject} and \emph{Accept} buttons on cookie banners. }    \label{fig:Banner_buttons}
\end{figure}

We asked respondents if they had ever encountered a cookie banner and presented them with a screenshot of one as an example. 
Most respondents (91\%) confirmed awareness, while 3\% responded no, and 6\% did not pay attention.
We also asked respondents about their experiences with the ease of finding and clicking on the \emph{Accept} and \emph{Reject} buttons on cookie banners, shown in~\autoref{fig:Banner_buttons}. Nearly half reported that the \emph{Reject} button was difficult to find (37\% difficult, 13\% very difficult), compared to only 21\% for the \emph{Accept} button (16\% difficult, 5\% very difficult). This highlights a significant usability disparity between the two options.

To further assess respondents' perceived interactions with different categories of cookie banners, we presented them with screenshots of four banners belonging to three categories in our survey (see~\autoref{fig:b1}, \autoref{fig:b2}, \autoref{fig:b3}, and \autoref{fig:b4} in the Appendix). We categorized these banners based on the work of Degeling et al.~\cite{degeling2018we}, which identified various implementations of cookie notices across websites. The categorization considered how the banners informed users about cookie usage, the type of consent mechanism used, and whether users \emph{could} manage or reject cookies. The three banner categories are:

\begin{enumerate}[noitemsep, topsep=0pt]
    \item Confirmation only (B1.1, B1.2): These banners included only an \emph{Accept} button.\footnote{While both B1.1 and B1.2 were of the same banner type, three factors differentiated them: (i) B1.2 could be ignored or closed, whereas B1.1 could not; (ii) B1.1 and B1.2 were displayed on different types of websites; and (iii) B1.1 included a detailed disclaimer on the banner, whereas B1.2 featured only a single-phrase disclaimer.}
    \item Binary (B2): This banner included both \emph{Accept} and \emph{Reject} buttons.
    \item No option (B3): This banner lacked both an exclusive \emph{Accept} and \emph{Reject} button.
\end{enumerate}

\autoref{tab:banners} shows the choices of interaction of respondents across various categories of cookie banners based on the available buttons on them.
Overall, when respondents had the option to \emph{Reject} a cookie banner (\textbf{B2: Binary}), nearly 45\% opted for the same. Still, a significant portion of respondents (about 34\%) also considered directly closing the same banner. 
When respondents were not given a choice to \emph{Reject} or \emph{Ignore} a cookie banner (\textbf{B1.1: Confirmation only}), their most preferred way of interacting was to click on \emph{Accept} (46\%), followed by an inclination towards exiting the website (30\%) and clicking on the \emph{Learn More} button (23\%). 
When the cookie banner lacked a \emph{Reject} button but did not interrupt a user's browsing experience (\textbf{B1.2: Confirmation only}), the most popular choice was to close the banner (41.5\%). But for the remaining respondents, a slightly higher fraction of them chose to click on \emph{Accept} (30\%) rather than ignoring the same banner (28\%).
For the case where respondents did not have an option to explicitly click on an \emph{Accept} or \emph{Reject} button (\textbf{B3: No option}), more than 46\% participants chose to avoid visiting such a website, followed by around 31\% choosing to exit the website. 
These findings show that the respondents interacted differently with different categories of cookie banners presented to them. Hence, the category of cookie banner (based on the available buttons) is a possible factor influencing the respondents' choice of interaction, an effect also noted in previous works~\cite{berens2022cookie, mathur2019dark}.

Note that banners B1.1 and B1.2 share the same category, yet respondents interacted differently with them. Although we did not specifically emphasize website types or disclaimer notices, they were visible in the screenshots. B1.1 could not be ignored or closed, and included a detailed disclaimer listing third parties. B1.2, however, could be ignored or closed, and a brief disclaimer was provided about enhancing browsing. Additionally, both appeared on different types of websites.
Thus, the option to ignore or close the banner, the type of website on which the banner is displayed, and the disclaimer notices could also be contributing factors influencing the respondents' decisions while interacting with the cookie banners. 
Previous work on users' interaction with cookie banners has also highlighted these factors~\cite{doi:10.1080/00913367.2017.1339368, kulyk2018website,schermer2014crisis}.  





\vspace{2mm}

\begin{mdframed}[linecolor=cyan!60!black, backgroundcolor=cyan!5]
\textbf{Key takeaway:} Most respondents confirmed their awareness of cookie banners. However, nearly half reported difficulty finding the \emph{Reject} option compared to the \emph{Accept} option, highlighting a design bias in cookie banners.
In addition, various factors influence users’ decisions when interacting with cookie banners, including banner category, type of website, the ability to ignore or close the banner, and the content of disclaimer notices.
\end{mdframed}

\subsection{RQ2: To what extent do Indian users value browsing privacy, and how do their expressed concerns align with their reported consent actions?}
\label{subsec:rq2}

We next analyzed users' broader perceptions of privacy. This helped us identify discrepancies between their privacy-related beliefs and choices, revealing potential gaps that could contribute to the privacy paradox \cite{gerber2018explaining}.
We presented the respondents with privacy-related statements (see Group 3 in~\autoref{group3}) and asked for their agreement on a 4-point Likert scale. 

\begin{figure} [h!]
    \centering    \includegraphics[width=\linewidth]{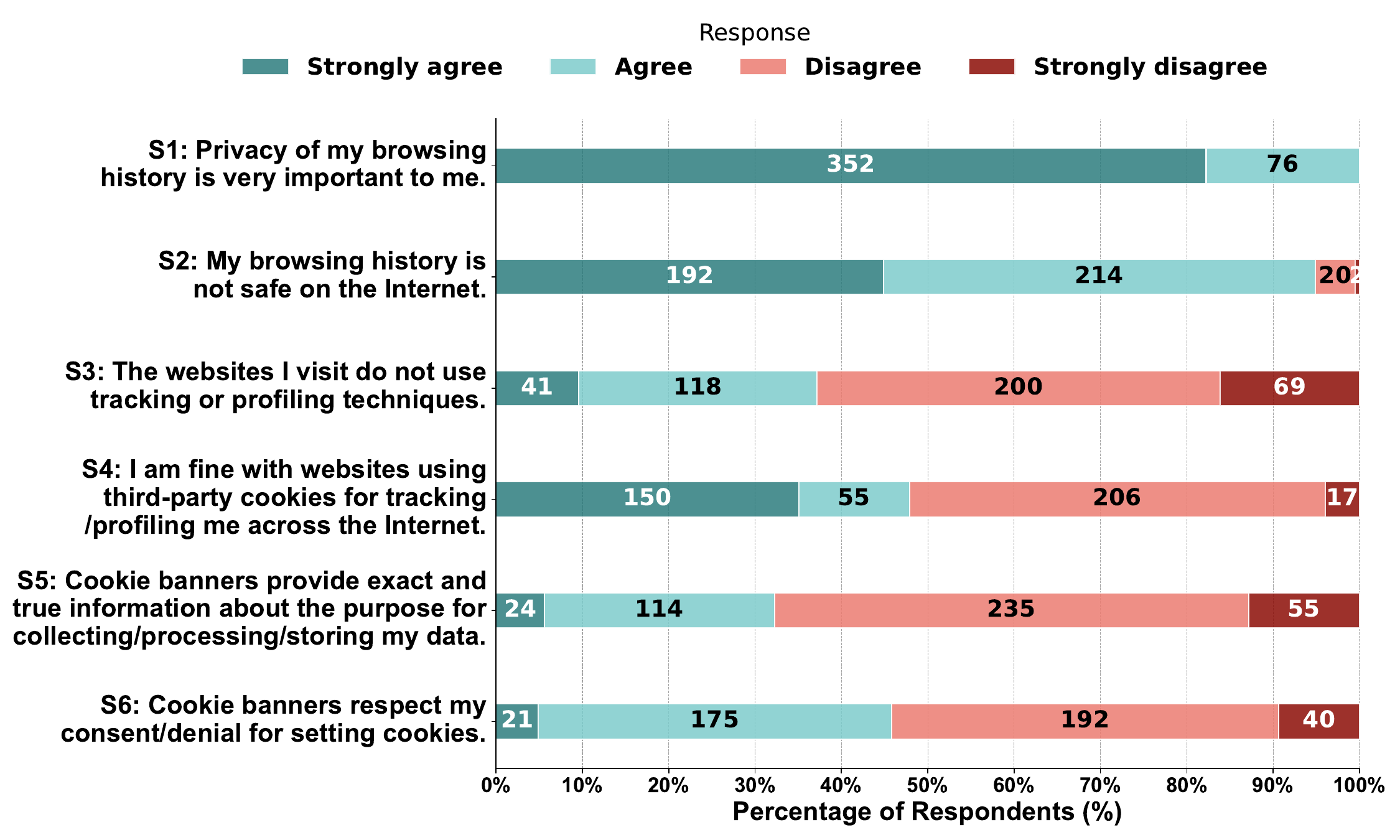}
    \caption{Users' perception of online privacy across browsing history, third-party cookies, and cookie banners.}
    \label{fig:fig5}
\end{figure}

\begin{figure*}[t!]
    \centering
    \begin{minipage}[b]{0.32\linewidth}
        \centering
        \includegraphics[width=\linewidth]{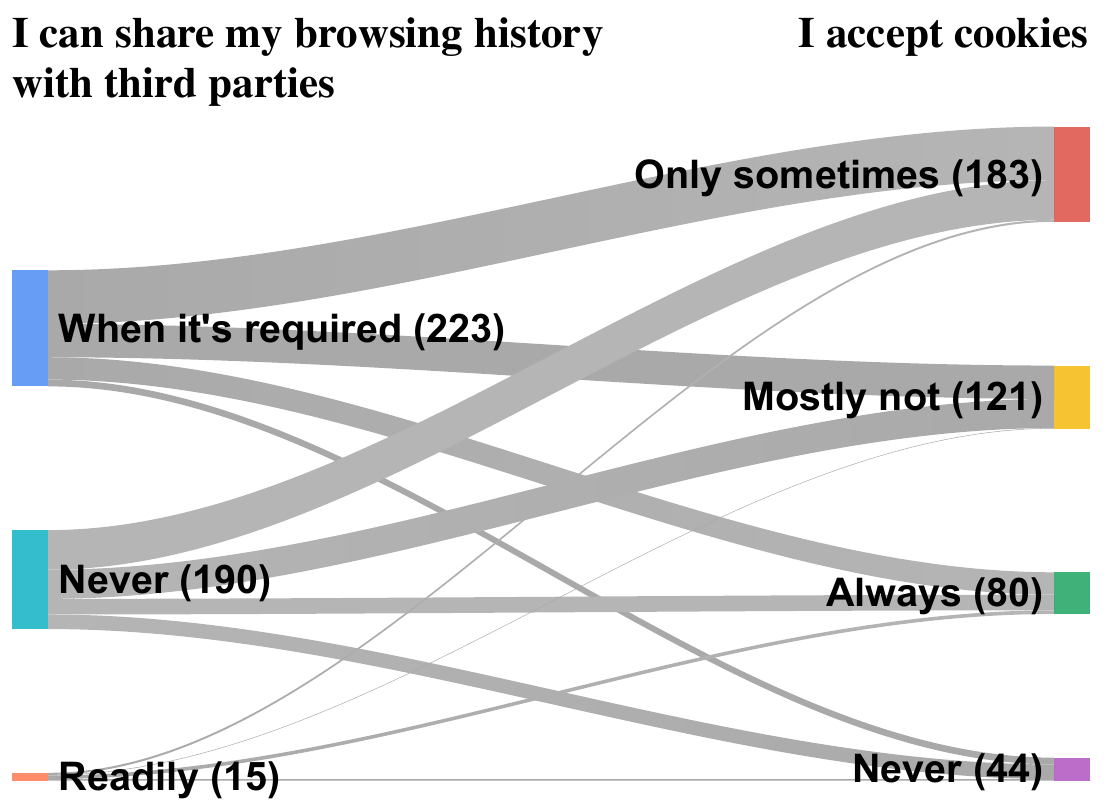}
        \caption{Browsing history sharing preference vs. cookie acceptance. }
        \label{fig:fig2}
    \end{minipage}
    \hfill
    \begin{minipage}[b]{0.32\linewidth}
        \centering
        \includegraphics[width=\linewidth]{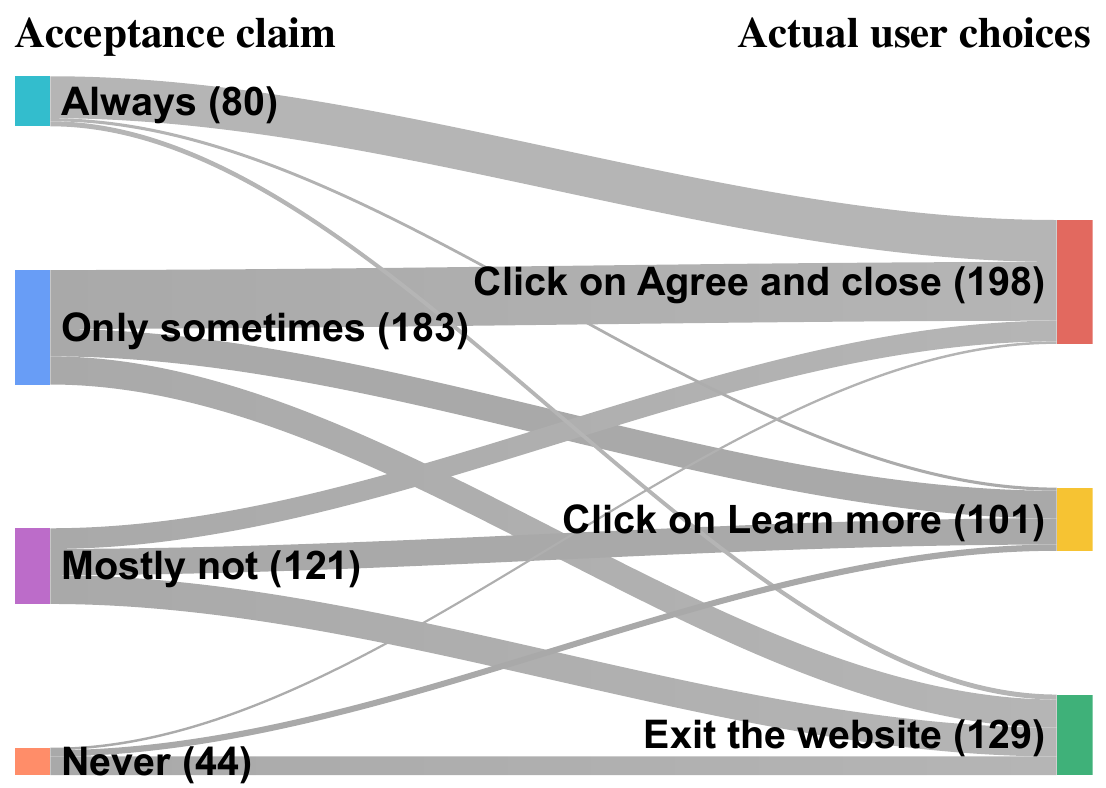}
        \caption{Cookie acceptance claim vs actual user choices.}
        \label{fig:fig3}
    \end{minipage}
    \hfill
    \begin{minipage}[b]{0.32\linewidth}
        \centering
        \includegraphics[width=\linewidth]{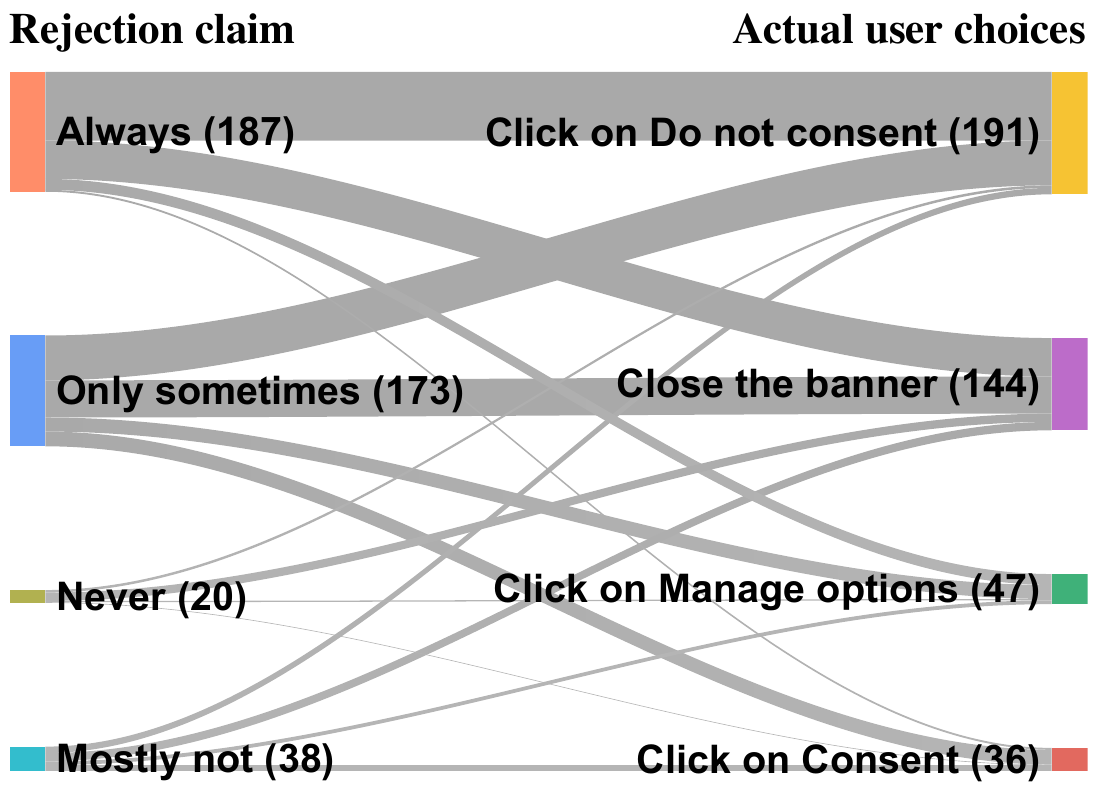}
        \caption{Cookie rejection claim vs actual user choices.}
        \label{fig:fig4}
    \end{minipage}
    
\end{figure*}

\noindent \textit{\textbf{Online privacy perceptions:}} To understand users' perception of the online privacy landscape, we evaluated the participants' responses for six privacy-related statements (see G03-Q.1 to G03-Q.6 in ~\autoref{group3}). The statements were related to browsing history, third-party cookies, and cookie banners (see~\autoref{fig:fig5}).

\begin{itemize}
    \item \textit{Browsing history:} Responses related to statements on browsing history (S1 and S2). 
    Participants highly value browsing privacy, with 94\% believing their history is unsafe online, reflecting widespread distrust in platform data handling.

    \item \textit{Third-party cookies:}  Responses related to statements on third-party cookies (S3 and S4).
    At least 37\% of respondents believe websites do \emph{not} use tracking or profiling, indicating a significant knowledge gap and potential misinformation about online tracking.
    Interestingly, a similar fraction of respondents express comfort (47\%) and discomfort (52\%) with websites tracking them across the Internet, indicating that many accept tracking and profiling.

    \item \textit{Cookie banners:} Responses related to statements on cookie banners (S5 and S6).  
    Nearly 68\% of respondents believe cookie banners do \emph{not} provide clear information on cookie purposes, suggesting awareness of dark patterns.
    Also, 54\% of respondents believe banners do \emph{not} respect their consent, highlighting significant trust issues in cookie compliance.
\end{itemize} 

\noindent \textit{\textbf{Browsing history sharing vs. cookie acceptance claim:}} After analyzing the users' perceptions of online privacy,
we analyze users' actual choices in realistic scenarios, comparing beliefs and actions to determine whether they align or reflect the privacy paradox, where attitudes and behaviors diverge.

We analyzed whether users' readiness towards sharing their browsing history with third parties aligns with their tendency to click on \emph{Accept} buttons on cookie banners. For this, we compare the responses between two questions from Group 1 and 2 (see questions G01-Q.8 
and G02-Q.2 in \autoref{questionnaire}), illustrated in ~\autoref{fig:fig2}.
Ideally, respondents who claim to \emph{never} share their browsing history with third parties would also be expected to avoid \emph{always} accepting cookies, as this enables third parties to track their activity across websites. However, our analysis reveals that out of 190 respondents who reported that they would \emph{never} share their browsing history with third parties, 30 claimed they \emph{always} accept cookies. Similarly, among the 223 respondents who stated they share their browsing history with third parties ``only when required,'' 44 also claimed they \emph{always} accept cookies, underscoring contradictions and the privacy paradox.

In addition, we conducted a Chi-squared test \cite{mood_statistics} ($\alpha = 0.01$) to examine the relationship between respondents' preference for sharing browsing history with third parties and how often they accept cookies. The analysis revealed no significant relationship ($\chi^2(6)= 8.50$, $p = 0.204$)  between these variables. This indicates that users' preferences for sharing browsing history with third parties do not significantly impact their decision to accept website cookies. Users also seem unaware of how cookies could be exploited by third parties to gain access to their browsing history. 

\noindent \textit{\textbf{Cookie acceptance claim vs. choice:}}
We examined whether respondents' claims about accepting cookies aligned with their actions of clicking the \emph{Accept} button on cookie banners. To assess this, we compared responses from two questions in Group 2 (see questions G02-Q.2 and G02-Q.3 in \autoref{group2}), shown in~\autoref{fig:fig3}. Out of 80 respondents who claimed that they \emph{always} accept cookies, 67 opted for clicking on \emph{Agree and Close}, being true to their claim. A Chi-squared test revealed a statistically significant relationship between the claim and choice of respondents regarding accepting cookies ($\chi^2(6)=100.72$, $p<0.01$), supporting the finding above.

However, there were notable instances of respondents contradicting their own claims.
Ideally, respondents who claim to \emph{only sometimes} accept cookies would be expected to click on the \emph{Learn More} and avoid selecting \emph{Agree and Close}. However, among 183 respondents who claimed to \emph{only sometimes} accept cookies, 94 chose to click on \emph{Agree and Close}, while only 44 opted for  \emph{Learn More}. The remaining 45 chose to exit the website. Moreover, out of 44 people who claimed to \emph{never} accept cookies, four still opted to click on \emph{Agree and Close}. 
These findings further highlight the contradiction between respondents' stated intentions and their reported cookie acceptance action.

\noindent \textit{\textbf{Cookie rejection claim vs. choice:}}
We analyzed whether respondents' reported claims about rejecting cookies translated into the action of clicking the \emph{Reject} button on cookie banners. 
To evaluate this, we compared the responses between two questions from Group 2 (see questions G02-Q.7 and G02-Q.8 in \autoref{group2}), depicted in~\autoref{fig:fig4}. 
We observed that out of 197 respondents who claimed to \emph{always}  reject cookies, 107 opted for clicking on ``Do not consent,'' aligning with their claim. A Chi-squared test again revealed a statistically significant relationship between the claim and choice of respondents regarding rejecting cookies ($\chi^2(9)=29.84$, $p<0.01$), validating the finding mentioned above. Hence, a user's decision to reject cookies was not independent of their claim, but we also observed instances showing a privacy paradox.
Ideally, respondents who claim to \emph{always} reject cookies would be expected to click on the ``Do not consent'' button when given the option. However, out of 197 respondents who claimed the same, 60 chose to ``Close the banner'' when they had the option to opt for ``Do not consent''.  Moreover, 13 of these respondents even chose to click the ``Consent'' button, completely contradicting their previous claim.  ``Close the banner'' has been a popular choice among respondents, regardless of their prior claims about rejecting cookies. This behavior suggests that users may prefer dismissing the banner over explicitly engaging with the consent options, likely due to misconceptions about consent options, unawareness, or privacy fatigue. 
This also shows the presence of a privacy paradox.

Our findings reveal gaps between respondents' privacy claims and choices, illustrating the privacy paradox. While they express privacy concerns, their actions often lack protective behaviors. As noted in \cite{gerber2018explaining}, this paradox stems from psychological, contextual, and informational factors, including unawareness, misinformation, trust, perceived benefits, cognitive biases, and privacy fatigue. Addressing this gap requires a multifaceted approach.

\vspace{2mm}
\begin{mdframed}[linecolor=cyan!60!black, backgroundcolor=cyan!5]
\textbf{Key takeaway:}
Indian users express concerns about online privacy and distrust in platform data handling, yet many remain unaware of third-party tracking, revealing a major knowledge gap.
Also, a clear privacy paradox emerges, as users' actions often contradict their stated intentions regarding cookie acceptance and rejection. 
Despite prioritizing privacy, users act out of unawareness, fatigue, or cognitive biases, highlighting the need to address psychological and informational factors to bridge the gap between privacy attitudes and behaviors. 
\end{mdframed}


\subsection{RQ3: What is the acceptance level of Indian users towards different clauses of DPDPA?}
\label{subsec:rq3}

We now present the acceptance levels of respondents towards the 10 clauses 
of DPDPA that we selected as they can lead to privacy violations \textit{en masse}  (see Group 4 clauses in \autoref{group4}).

\begin{figure}[t!]
    \centering
    \includegraphics[width=0.45\textwidth]{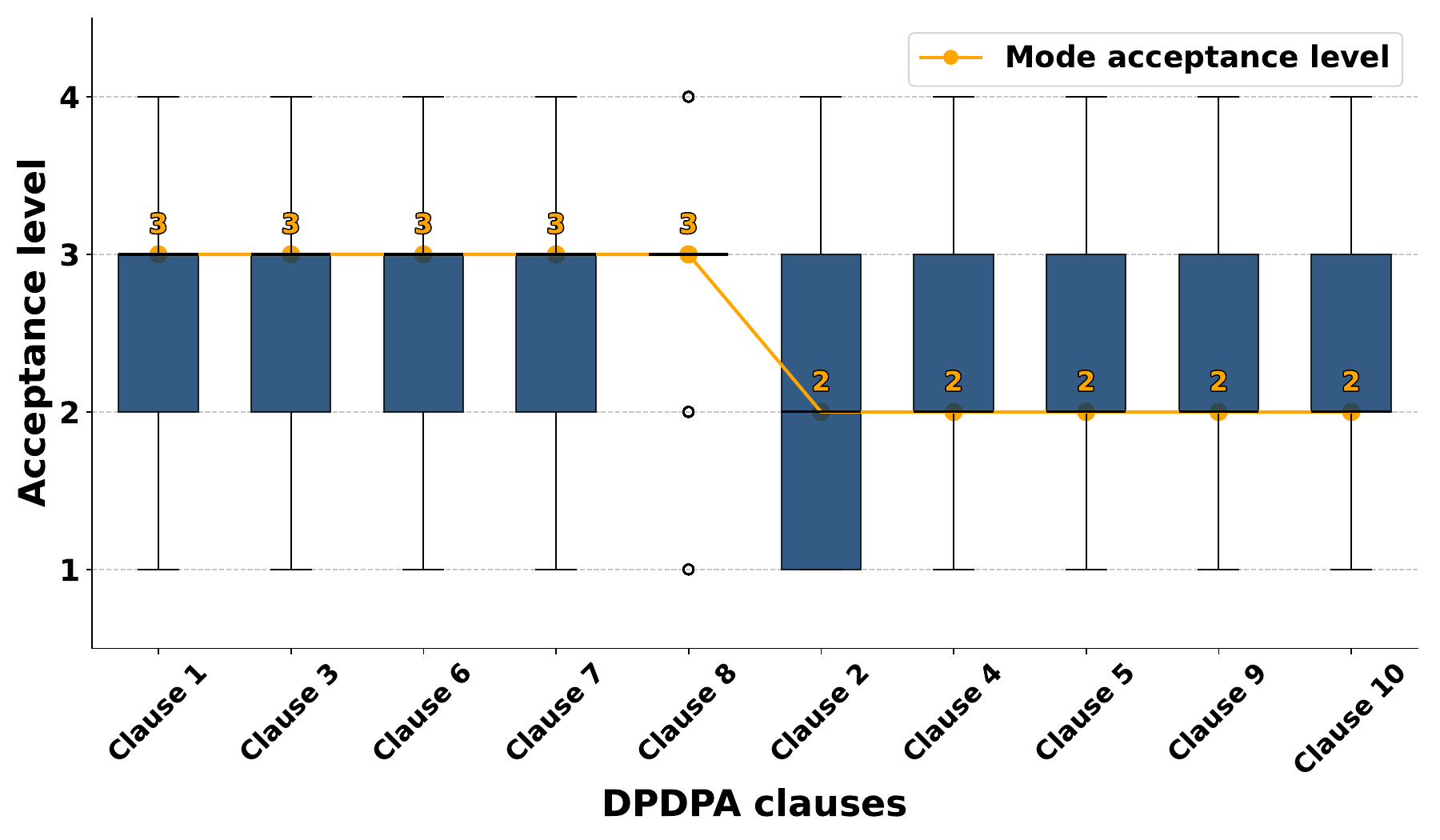}
    \caption{
    Acceptance levels of 10 DPDPA clauses linked to potential privacy violations. Mode values represent most preferred reported acceptance from 1 (highly unacceptable) to 4 (highly acceptable).}
    
    \label{fig:fig6}
\end{figure}

To determine if there were significant differences in the acceptability of these clauses, we performed a Kruskal Wallis test (H0: There are no significant differences in the acceptance level among the 10 clauses, $\alpha = 0.01$). The null hypothesis was rejected (\textit{H(9)=316.04, p<0.01}), establishing that there are statistically significant differences in the acceptance levels. To obtain a better understanding of the general consensus of respondents towards each clause, we also determined mode acceptance levels using the ranks previously assigned to the acceptance levels (as discussed in ~\autoref{subsec:analysis}). ~\autoref{fig:fig6} shows the acceptance levels of all clauses. The \emph{mode} for every clause indicates the mode acceptance level of that clause on a scale of 1 to 4 (highly unacceptable, unacceptable, acceptable, and highly acceptable).

Clauses 1, 3, 6, 7 and 8 have mode value 3, suggesting they are viewed more positively. This indicates that Indian citizens are relatively more comfortable with DPDPA clauses stating that 
(i) the government can process their personal data without their consent if it's in the interest of the security and integrity of the nation (Clause 1), (ii) the tenure of the Data Protection Board of India is two years with scope for re-appointment (Clause 6) and (iii) the DPDPA does not apply when individuals voluntarily share personal data, such as blogging their views online (Clause 8).

Clauses 2, 4, 5, 9, and 10 have mode value 2, indicating that they are unacceptable to most of the respondents. 
These clauses state that (i) the Government may not require users' consent for processing their personal data for specified \textit{legitimate purposes} (Clause 2), (ii) no lawsuit can be filed against the Government, the board or any of its members for anything which they do in \textit{good faith} (Clause 9) and (iii) the act allows transfer of personal data outside India for \textit{legitimate purposes} unless restricted by the Government through a notification (Clause 10). We hypothesize that the low acceptance levels for clauses 2, 9, and 10 are due to the presence of vague terms like \textit{legitimate purposes} and \textit{good faith}, which are undefined in the law.
Overall, possible explanations for acceptance levels of these clauses are further discussed in thematic analysis (\autoref{t2}).




\vspace{2mm}
\begin{mdframed}[linecolor=cyan!60!black, backgroundcolor=cyan!5]
\textbf{Key takeaway:}
Indian users' acceptance of DPDPA clauses varies significantly, with higher acceptance for clauses related to national security, data protection board tenure, and voluntary data sharing, but lower acceptance for clauses allowing government data processing without consent, immunity from lawsuits, and cross-border data transfers, highlighting concerns about accountability and data privacy.
\end{mdframed}

\subsection{Thematic analysis: Open-ended responses}
\label{subsec:openfield}
The optional open-ended field at the end of the survey received 143 responses. We qualitatively analyzed these using thematic analysis~\cite{clarke2017thematic} and generated three main themes.
\subsubsection{T1: User concerns about privacy, data protection, and cookies} \label{t1}
This theme encapsulates users' concerns about online privacy---including cookies---and data protection, as well as their emphasis on users' autonomy and the need for stringent data protection regulations. A notable number of entries (n=14) expressed the importance of online privacy and the need for data protection. For instance,
\begin{quote}
    \textit{``Data protection is essential in the digital era to safeguard personal and sensitive information from breaches.''}
\end{quote}
\begin{quote}
    \textit{``Protection of personal data is, directly or indirectly, related to the right to privacy and Govt. should ensure that it is protected with provisions in the law to safeguarding it.''}
\end{quote}

Some participants expressed concerns about the current data privacy situation (n=4), stating, e.g., that \textit{``every gadget one uses is mining personal data''}. Others mentioned the grave consequences of personal data leakage (n=3) for national security. To mitigate these concerns, data protection regulation (n=3) is called upon as the \textit{``need of the hour''}.

Additionally, respondents highlighted the importance of user consent (n=5) and autonomy (n=1). They state that, 
\begin{quote}
    \textit{``No one should access anyone's personal data without individual permission.''}
\end{quote}
\begin{quote}
    \textit{``Nobody should control end user.''}
\end{quote}

Many raised issues regarding how cookie banners are handled by websites (n=5), with one stating that,
\begin{quote}
    \textit{``In general, common people who are browsing the Internet don't know what cookies are. They simply accept because they do not find reject or decline options. 
    Moreover, on many websites, you cannot proceed further if you don't accept cookies.''}
\end{quote}
The aforementioned quote also highlights other aspects such as lack of digital literacy among the general population, a point highlighted by two respondents, and the lack of rejection options on cookie banners.
Two entries call for the incorporation of equal and fair reject buttons, with one stating that the \textit{``default setting should be reject all''}.
Respondents perceived cookies as problematic, attributing them as an attack surface for data leakage incidents and claiming that cookie banners don’t provide complete and true information. 


\begin{quote}
    \textit{``Lot of times cookies collect data which is not approved or approval is hidden behind twisted wording.''}
\end{quote}

This theme uncovers additional factors for the respondents' reported choice of interactions with the cookie banners, providing further context to our findings for \textbf{RQ1} (see~\autoref{subsec:rq1}). 
We observe that there is sensitivity yet unawareness about online privacy, which plays a major role in influencing users' decisions. 
The lack of \emph{Reject} buttons on cookie banners, with many websites making it difficult to browse through without interacting with the banners, forces users to accept the cookies. 
\subsubsection{T2: Improving the DPDPA framework} \label{t2} 
The core idea of this theme is respondents' views on how the current DPDPA framework can be enhanced. 
DPDPA is felt to still be in a \textit{``nascent stage''} (n=2), requiring significant revisions (n=6) and \textit{``more clarifications''} (n=2). 
Respondents state that,
\setlength{\topsep}{0pt}
\setlength{\partopsep}{0pt}
\setlength{\parskip}{0pt}
\setlength{\parsep}{0pt}
\setlength{\leftmargin}{-5pt}%
\setlength{\rightmargin}{0pt}%
\begin{quote}
    \textit{``It needs to be relooked upon by the parliament (both houses).''}
\end{quote}
\begin{quote}
    \textit{``Th[e] act is at least a decade behind in its current form and needs amendments and updates asap.''}
\end{quote}

Suggestions for improvement include modification in chairperson appointment procedure (n=3), with one participant elaborating as follows:
\begin{quote}
    \textit{``Chairperson should not be confined to the central government, but various stakeholders like the leader of opposition in Lok Sabha\footnote{Lok Sabha is the lower house of India's bicameral Parliament, composed of representatives of people chosen by direct election.}, CJI\footnote{Chief Justice of India.} and any other stakeholders should be made so, there seems some loopholes which may generate an arbitrariness in the appointment as well as extension of period of term of the chairperson.''}
\end{quote}

Respondents also suggest the need for more stakeholder involvement (n=3), e.g., by enacting the law \textit{``after an open debate with all lawmakers in the central and state assemblies''} and for more transparency (n=2) both by \textit{``government and companies''} when one's data is \textit{``shared with third parties''}.
Two respondents advised learning from other countries, with one remarking that \textit{``Indian policymakers have learned almost nothing from Europe and the United States''}.

There is also concern about government exemptions from these policies.
\begin{quote}
    \textit{``Government should not make any kind of exemption for any authority, organization or individual.''}
\end{quote}
While some call for no exemptions (n=3), others accept these exemptions regarding national security and integrity (n=4). For instance,
\begin{quote}
    \textit{``The government makes a law that protects our digital privacy, and even the government does not violate those rights unless it's a national emergency.''}
\end{quote}
These inputs from respondents corroborate the higher acceptance of Clause 1 by the respondents, providing an explanation for our findings for \textbf{RQ3} (see~\autoref{subsec:rq3}).
While some codes portray a constructive critique of DPDPA, some respondents show substantial pessimism towards it (n=8). For instance,

\begin{quote}
    \textit{``New provisions coming into force in DPDP Act, totally nullifies the whole idea of Right to Data privacy.}''
\end{quote}

\begin{quote}
    \textit{``There is effectively no improvement in privacy and no recourse for Indian citizens.''}
\end{quote}


\begin{quote}
    \textit{``DPDP Act in itself is an incomplete and non-comprehensive legislation.''}
\end{quote}

\begin{quote}
    \textit{``Dangerous and will lead to a police state.}''
\end{quote}

These responses could explain the lower acceptance of Clauses 2, 9, and 10 (see~\autoref{subsec:rq3}). 
There are, however, instances where respondents have shown faith in the existing state of DPDPA, emphasizing that it caters to users and businesses (n=2) by \textit{``addressing longstanding needs in the context of increasing Internet users, data generation, and cross-border trade''} and that \textit{``it mandates a significant shift from how Indian businesses should now approach privacy and Personal Data, while legitimizing Central Government to control, retain, and monitor its citizens’ personal information''.}
There are also entries that express that the DPDP Act promotes privacy (n=2) and
that should be implemented as it is (n=2):

\begin{quote}
    \textit{``The Act marks a distinctive approach to safeguarding Personal Data.''}
\end{quote}

\begin{quote}
    \textit{``Such act, rule, regulation must be adopted for Indian citizens, which can provide data security to them.''}
\end{quote}

Overall, both from the quantitative and qualitative findings, we observe a partial acceptance for DPDPA, a sentiment echoed
by one of the responses, 

\begin{quote}
    \textit{``I am fine with this act, but at the same time, I do not agree with some articles of this act.''}
\end{quote}

Lastly, some responses were directed towards unawareness about DPDPA (n=3) and the online privacy landscape in general (n=3):

\begin{quote}
    \textit{``I don't [have] the full knowledge of DPDP.''}
\end{quote}
\begin{quote}
    \textit{``[Necessary] to spread awareness among citizens through surveys like this where we find out more about the protection of our data online.''}
\end{quote}
These responses mentioned above point to the larger problem of digital literacy about online privacy among the general Indian population and the need for a coordinated effort to spread awareness to combat this.

\subsubsection{T3: Trust-distrust in the system} \label{t3}
This theme covers the various levels of trust and distrust of respondents towards the people in the system, mainly the Government. 
Respondents expressed trust in the government (n=1) and judiciary (n=1) of the country by stating:
\begin{quote}
    \textit{``[...] as we are citizens of India now only can trust supreme courts bench appointed authority and rules and guidelines.''}
\end{quote}
\begin{quote}
\textit{"Bhartia/Indian govt. is moving in the right direction."}
\end{quote} 

We also find support for nationalism (n=1) and authoritarianism (n=1):

\begin{quote}
    \textit{``Left ideology uses these data to benefit a certain group while nationalist govts use for the nation.''}
\end{quote}
\begin{quote}
    \textit{``I love authoritarian regimes, this is what has been decreed, respect the authority and do not rebel, otherwise you're a westernized emotional fool who likes the idea of "free speech".''}
\end{quote}


On the other hand, respondents have shown two types of distrust towards the Government, with regards to (a) their intentions regarding the collection and processing of user data (n=6) and (b) their capability to protect user data (n=2). 

\begin{quote}
    \textit{``Government will use our information [and] data as and when required without our consent.''}
\end{quote}

\begin{quote}
    \textit{``There have been multiple instances of government data breaches in the past.''}
\end{quote}

Respondents believe that government surveillance and incompetence in handling user data responsibly pose significant threats to the privacy of the citizens (n=5). For instance, 

\begin{quote}
    \textit{``Government spying on it's citizens is not acceptable.''}
\end{quote}

\begin{quote}
    \textit{``Government should not just implement the things it should also keep check back mechanism.''}
\end{quote}

\begin{quote}
    \textit{``If not government then who will take the responsibility?''}
\end{quote}

This theme portrays the contrast between trust and distrust, and the perceived lack of accountability in governmental bodies regarding data protection. It shows a desire for more responsible handling of user data but reflects a deep skepticism about whether the government can fulfill this role without infringing on the personal privacy of users keeping in view the exemptions (see the clauses in Group 4 in~\autoref{group4}).

\subsubsection{Interrelatedness}
\label{subsubsec:themerelations}
The three themes are not isolated but rather form a cohesive narrative about respondents' concerns.
\textit{T2: Improving the DPDPA framework} emerges as a response to \textit{T1: User concerns about privacy, data protection and cookies}, while both of these are driven by \textit{T3: Trust-distrust in the system}. 

\begin{figure}[h]
    \centering
    \includegraphics[width=0.6\linewidth]{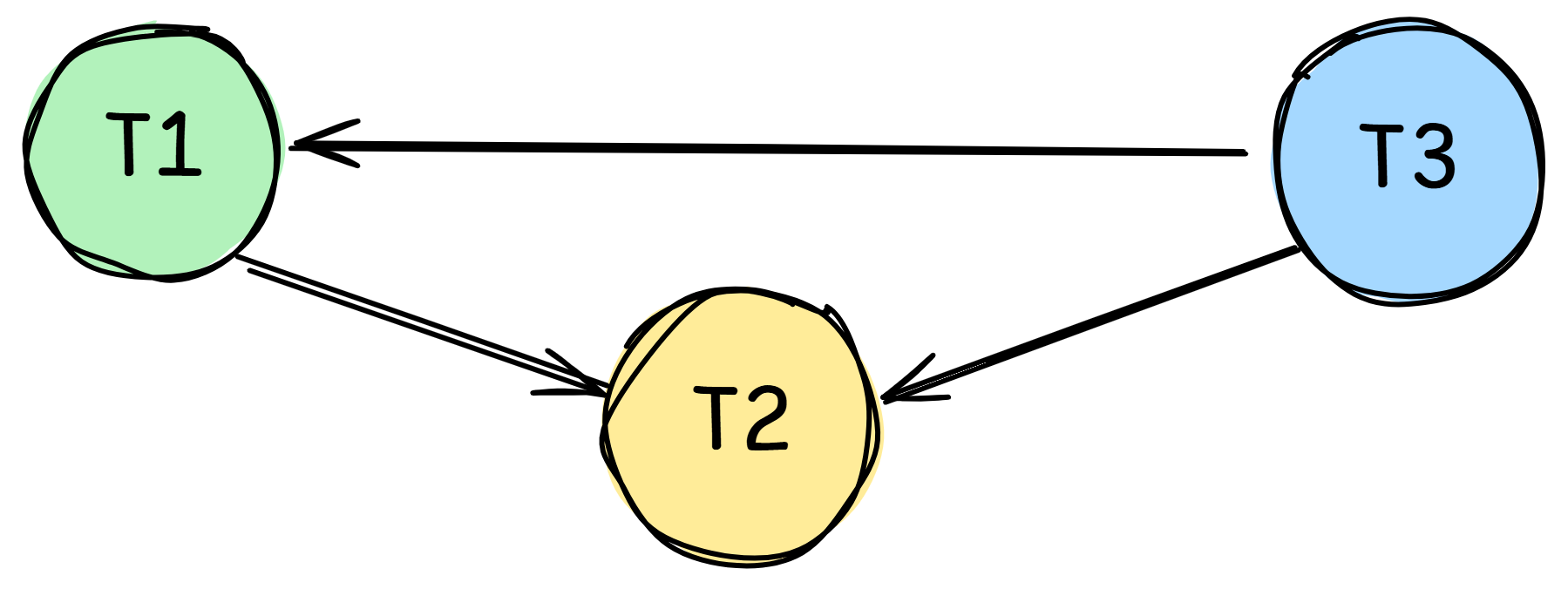}
    \caption{The three themes (T1, T2 and T3) are inter-related.}
    \label{fig:inter-related}
\end{figure}

The first theme, \textit{T1}, leads to the second theme, \textit{T2}. Where the second theme emphasizes the need for structural improvements in the DPDPA, the first theme captures the driving force behind these demands: a deep-seated concern about online privacy violations and data insecurity. The respondents' suggestions to improve the DPDPA framework stem also from their belief that the current regulatory landscape is inadequate in addressing these privacy concerns---connecting \textit{T2} to \textit{T3}. This distrust informs many of the specific suggestions for DPDPA improvements. The push for removing Government exemptions in the DPDPA stems from a belief that the Government could misuse these exemptions to infringe upon personal privacy, which is reflected in \textit{T3}. Similarly, \textit{T1} and \textit{T3} represent two sides of the same concern. While \textit{T1} focuses on the specific anxieties users feel about the safety of their personal information, \textit{T3} delves into the source of these anxieties: the distrust in the Government’s ability to handle data properly. For many respondents, the perceived risks to their personal data are rooted in a broader distrust of the Government’s motives and capabilities, and both \textit{T1} and \textit{T3} emphasize the need for greater accountability.

\vspace{2mm}
\begin{mdframed}[linecolor=cyan!60!black, backgroundcolor=cyan!5]
\textbf{Key Takeaway:} Respondents express significant concerns about online privacy, data protection, and cookies (T1), emphasizing the need for structural improvements in the DPDPA framework (T2) while highlighting distrust in the government’s ability to handle data responsibly (T3). These themes are interrelated, with privacy concerns driving demands for regulatory improvements, which are shaped by skepticism about government motives and accountability. 
\end{mdframed}

\section{Discussion}
\label{sec:discussion}


\subsection{Key insights compared to prior work}

\noindent \textbf{Various factors influence user interactions with cookie banners:} Prior works have identified banner design, placement, disclaimer content, button labels, trust in the website, ease of rejection, and desensitization \cite{doi:10.1080/00913367.2017.1339368, berens2022cookie, kulyk2018website, chanchary2015user, mathur2019dark} to be some common factors that influence users' engagement with cookie banners mostly from European context. 
In our study, we provide Indian users with screenshots of different types of actual cookie banners displayed on websites. This methodology helped us understand users' preferred way of interacting with banners, taking into account the entire context (banner placement, buttons, disclaimer notice, ease of ignoring/dismissing, and the website). We found that some factors identified by prior studies, such as ease of rejection \cite{mathur2019dark}, button labels \cite{berens2022cookie}, and website type \cite{kulyk2018website}, still hold in our study. However, Indian users more frequently cite uncertainty and a lack of meaningful choice as reasons for accepting cookies, reflecting the region’s minimal exposure to privacy regulations. Despite numerous concerns about browsing privacy, Indian participants exhibit relatively low engagement with banner customization--- possibly due to limited exposure to such interfaces and low digital literacy---a sentiment also reflected in our qualitative data. Additionally, we observed that \emph{closing the banner} has been a preferred choice (apart from \emph{Reject}) by notable respondents whenever they had the option to do so. This suggests that fatigue, confusion, misinformation, misconception, or irritability can cause users to disengage from choosing any consent option. 
\vspace{1mm}

\noindent \textbf{Online privacy knowledge and the privacy paradox:} Previous research has highlighted that convenience, unawareness, underestimation of risks, fatigue, cynicism, and cognitive effort \cite{barth2017privacy, norberg2007privacy, stutzman2013silent, hoffmann2016privacy, potzsch2009privacy} lead to privacy paradox---a phenomenon where users, despite being privacy-conscious, fail to engage in privacy-protective behaviors.
Our findings also confirm this: users are often unaware or misinformed about online privacy. Their claim to \emph{never} share browsing history with third parties contradicts their tendency to \emph{always} accept cookies, suggesting that either they are unaware of the association and implications of these actions or they are misguided. Additionally, users' claims about accepting and rejecting cookies often did not align with their choices. Instances where users first claimed to \emph{only sometimes} accept cookies, then opted for \emph{Agree and Close} despite having the \emph{Learn More} button, indicate that they are privacy conscious in their intentions, but don't necessarily take extra efforts to ensure that privacy in their actions.
However, the question of design bias remains: Why should users make this extra effort to ensure privacy?
\vspace{1mm}

\noindent \textbf{Acceptance levels of users towards privacy regulation}: While existing studies have explored users' opinions about GDPR \cite{mangini2020empirical, strycharz2020data, kyi2024doesn, gati2020perception}, DPDPA has not been studied in this context yet. Moreover, prior works have focused on user expectations and awareness of provisions and implementations in the context of GDPR \cite{mangini2020empirical, strycharz2020data}, but no study highlights user \emph{acceptance} of such privacy laws. Our study provides a novel insight into this domain, wherein users have expressed varying acceptance levels towards some contentious DPDPA clauses. If not implemented with strict oversight, these clauses could be exploitative and lead to surveillance. Our research provides timely insights that can help policymakers revisit and refine these clauses of DPDPA before its enforcement. By addressing public concerns and prioritizing user trust, they can ensure that the act achieves its primary goal of safeguarding user privacy effectively rather than compromising it in certain scenarios.
\vspace{1mm}

\noindent \textbf{Alignment with a parallel company survey}: Our study aligns with and complements broader national findings—such as the PwC-India\footnote{Indian branch of PwC (PricewaterhouseCoopers), a global network of firms providing assurance, tax, legal, deals, and consulting services.} survey \cite{PwC2024DPDPAsurvey}—in important ways. Their large-scale, demographically diverse survey (over 3,000 consumers across age groups, city tiers, and occupations) reveals a striking awareness gap: less than one-fifth of consumers know about DPDPA, and trust in organizational data practices remains low. In contrast, our participant pool—composed of educated, digitally active individuals—voluntarily engaged in the study and demonstrated greater familiarity with the Act and nuanced privacy perceptions. This positions our work as a purposeful exploration of the segment most likely to understand, respond to, and be impacted by privacy regulation. Whereas PwC’s survey examined privacy concerns across a wide range of contexts—including Aadhaar, email IDs, and PAN numbers—our study focused specifically on web privacy through users’ interactions with cookie banners. This narrower lens is particularly relevant, as one of the most visible manifestations of the DPDPA for everyday users is likely to be an increase in consent requests on websites.

\vspace{1mm}



\noindent \textbf{Interconnected concerns about data handling systems, privacy, and trust in regulation and policymakers:} Our interconnected themes (see~\autoref{subsubsec:themerelations}) highlight how users are anxious about the safety of their data online, which is rooted in their skepticism and distrust towards data processors and the policymakers. 
The qualitative responses highlight the call for improvements in the DPDPA, along with widespread skepticism about the intentions and capabilities of policymakers. 
This distrust is critical to address, as it undermines the effectiveness of the law. Previous studies on GDPR have also highlighted that users are often aware of their rights but confused about the scope to which they can exercise them \cite{mangini2020empirical}. Moreover, despite GDPR, many users still believe companies will misuse their data \cite{strycharz2020data}. If users lack confidence in policymakers, they are unlikely to feel secure or satisfied with privacy regulations. They may hesitate to exercise their rights under the law---ultimately defeating its primary purpose of safeguarding user privacy in a democratic society. 

\subsection{Authors' neutrality and clause selection}
Our study is grounded in the objective reporting of user perceptions and acceptance of the DPDPA, and does not aim to endorse or critique the regulation. As authors, we have maintained neutrality throughout and focused solely on presenting the views expressed by survey participants. The ten clauses included in the survey were selected based on their prominence in public discourse, media reports, and policy analyses---particularly those believed to have significant implications for user privacy, consent, and data protection. We recognize that additional clauses in the DPDPA may also be subject to ambiguity, misuse, or conflicting interpretation. However, our selection was guided by practical constraints and relevance to everyday user experience. Ultimately, our goal was to foreground public perceptions---accurate or otherwise---as these shape how individuals engage with privacy regulations and build trust (or distrust) in institutional frameworks.

\subsection{Limitations and future work}
\label{sec:limitations}


Our study uses static screenshots that cannot fully capture the dynamic and contextual nature of real-world browsing, and we do not claim this to be a study in the wild. Cookie banner interactions were chosen as a proxy for privacy attitudes due to their ubiquity on websites and their function as the first point of user-data negotiation. Unlike abstract questions about privacy concern, banner interactions offer an insightful lens into users’ consent decisions. Our use of actual banner screenshots—with placement, options, and website context—enhances ecological validity and helps ground privacy attitudes in realistic decision-making contexts.

These sample banner designs have also been highlighted \cite{degeling2018we} and utilized in previous works~\cite{kulyk2018website, bouma2023us}. 
Additionally, conducting an online survey on a self-hosted platform without financial incentivization, especially one as extensive as ours with 38 questions, proved challenging for garnering a high volume of responses.
Despite these hurdles, we obtained 428 complete responses.
While we do not claim generalizability, with our self-hosted anonymous survey tool and non-incentivized voluntary approach, we aimed to get input from a more diverse demographic than crowd-sourced platform workers. 
Moreover, a quantitative survey approach also meant that the responses were limited in depth and that there was a risk of random responses. 
Using an unpaid and anonymized survey mitigated this risk of obtaining random and dishonest responses.
Furthermore, we provided participants the option of sharing feedback and comments with us at the end of the survey, which garnered responses from around 33\% of the respondents (143/428). 
We qualitatively analyzed these inputs and reported them as part of our findings.
Lastly, our data exhibited some geographical bias.
Specifically, 52\% of the responses came from the state of Uttar Pradesh (UP).
However, isolating responses from UP showed that the proportion of chosen answers remained consistent and did not affect the overall trends observed in our findings.


For future work, we intend to explore longitudinal qualitative studies on Indian users' evolving attitudes as DPDPA is enforced and cookie banners become more prevalent. In addition, compare privacy perceptions in different regions of India, considering its socio-cultural diversity, or delve into sector-specific privacy concerns (e.g., healthcare data).

\section{Conclusion}
We surveyed 428 educated Indian participants and analyzed their reported interactions with cookie banners, attitudes toward online privacy, and acceptance levels regarding specific DPDPA clauses that raise privacy concerns. Our findings revealed that while Indian users are generally sensitive to privacy, they often lack a basic understanding of consent mechanisms and data protection regulations, indicating gaps in awareness that may affect their engagement with privacy tools. Notably, respondents expressed mixed acceptance of certain DPDPA clauses that allow government exemptions, with clauses on national security receiving higher acceptance than those restricting lawsuits against the government,
highlighting public concerns about transparency and accountability. 
Overall, this study provides a foundation for understanding user \emph{perspectives} in India’s privacy landscape, informing strategies that promote a balanced and trust-centric approach to data protection.

\bibliographystyle{ACM-Reference-Format}
\bibliography{references}
\appendix

\section{APPENDIX}
\subsection{Respondents' distribution across Indian states}
~\autoref{tab:states} illustrates the number of participants from different states of India who completed our survey. 
\label{subsec:states}

\begin{table}[htb!]
    \centering
    \resizebox{\linewidth}{!}{
    \begin{tabular}{|c|r||c|r|}\hline 

        \textbf{State} & \textbf{Respondents} & \textbf{State} & \textbf{Respondents} \\ \hline
        Uttar Pradesh & 225 & Kerala & 3 \\ \hline
        New Delhi & 54 & Madhya Pradesh & 3 \\ \hline
        Non-resident Indian & 45 & Tamil Nadu & 3 \\ \hline
        Karnataka & 25 & Gujarat & 2 \\ \hline
        Bihar & 21 & Jharkhand & 2 \\ \hline
        Maharashtra & 11 & Uttarakhand & 2 \\ \hline
        Haryana & 10 & Andhra Pradesh & 1 \\ \hline
        Assam & 6 & Chandigarh & 1 \\ \hline
        West Bengal & 5 & Himachal Pradesh & 1 \\ \hline
        Jammu and Kashmir & 4 & Nagaland & 1 \\ \hline
        Kerala & 3 & Punjab & 1 \\ \hline
        Madhya Pradesh & 3 & Rajasthan & 1 \\ \hline
        Tamil Nadu & 3 & Telangana & 1 \\ \hline
    \end{tabular}
    }    
    \caption{Respondents' distribution across states In India (n=428).}
    \label{tab:states}
\end{table}

\subsection{Privacy Notice}
\label{privacynotice}
This survey will take about 15 minutes to complete. In the next section, we will be collecting some of your personal information like Age, Gender, City, Professional Status, etc., which are not a direct identifier of your identity. We collect this information to report the demographic information as part of our study’s findings. After completing the survey, your responses will be reported in an aggregated/anonymized way. Participation in this study is entirely voluntary, and you can withdraw your participation at any time before you click on the ``Submit'' button. If you agree with these terms, kindly check the box below and head to the next section. 

\subsection{Survey Questionnaire}
\label{questionnaire}
\subsection*{\textbf{Group 1: About You (G01)}}
\label{group1}

Q.1 Age:
\begin{itemize}
    \item 18-25
    \item 26-35
    \item 36-45
    \item 46-55
    \item 56-65
    \item 66-75
    \item Above 75
\end{itemize}

Q.2 Gender: (Open-field) \newline
Q.3 State/Territory of residence: (\emph{Please choose Non-Resident Indian if you reside outside India} \newline 
(drop-down menu) \newline 
Q.4 City: (\emph{Please fill ``Other'' if you are a Non-Resident Indian")} \newline
(Open-field) \newline 
Q.5 Which of the following best describes your professional status? 
\begin{itemize}
    \item Student
    \item Academic Researcher
    \item Industry Professional 
    \item Other: (Open-field)
\end{itemize}
Q.6 What is your field of study/research/work? 
\begin{itemize}
    \item Computer Science
    \item Mathematics/Statistics/Physics/Chemistry or other Science related fields
    \item Engineering and related fields 
    \item Medical Sciences and related fields 
    \item Humanities/Law/Journalism and related fields 
    \item Commerce/Finance/Economics/Business/Management or other related fields 
    \item Other: (Open-field)

\end{itemize}
Q.7 How often do you use the Internet, e.g., for social media platforms, web browsing, or work?
\begin{itemize}
    \item Everyday
    \item 3-4 times a week
    \item Maybe once a week
    \item Less than once a week
    \item Other: (open-field)

\end{itemize}
Q.8 How comfortable are you with sharing your browsing history with advertisers/online tracking agencies? \label{G01Q08}
\begin{itemize}
    \item I can share readily
    \item I can share only when it's required
    \item I can never share it
\end{itemize}
Q.9 What is your preference regarding the privacy of your browsing history?
\begin{itemize}
    \item Extremely important to me that only the intended party accesses it
    \item I would feel good if only the intended party accesses it
    \item Doesn't bother me if a stranger sees my pictures and/or gets my number
    \item I've got nothing to hide
\end{itemize}

\noindent\hrulefill 

\subsection*{\textbf{Group 2: Your Experience (G02)}}
\label{group2}
\begin{figure}
    \centering
    \includegraphics[width=0.75\linewidth]{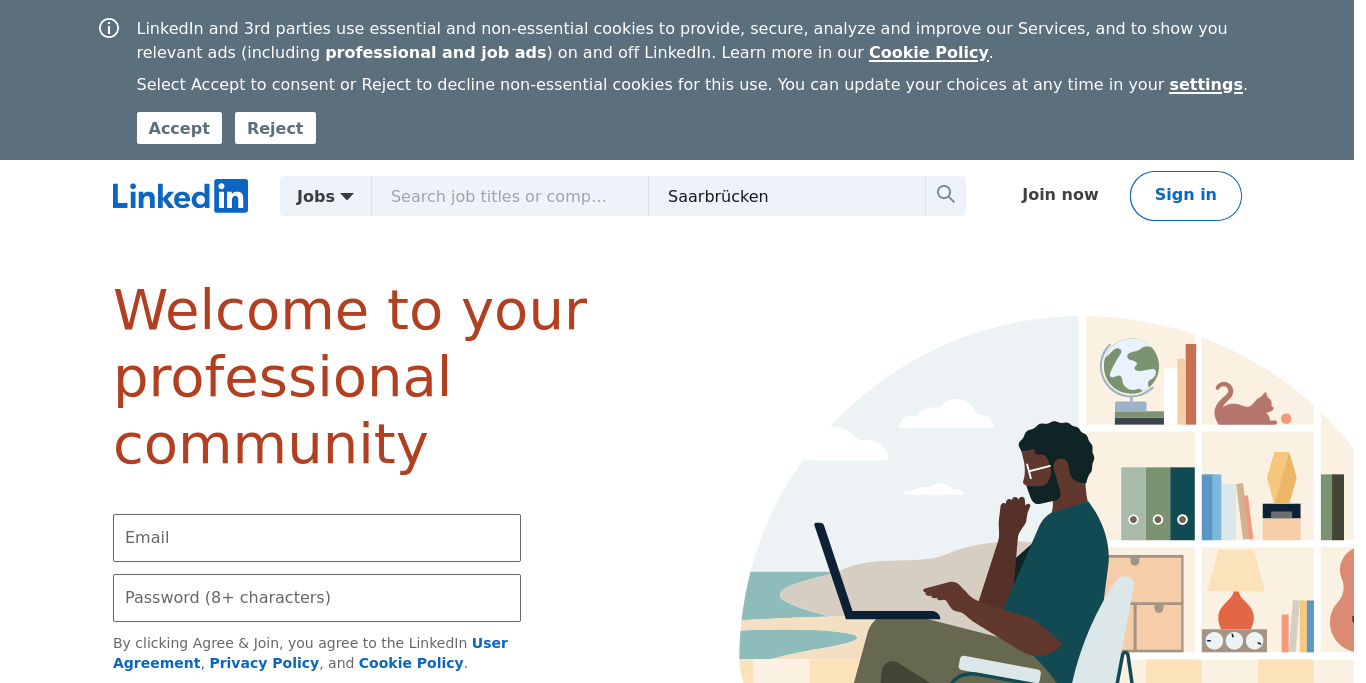}
    \caption{Banner screenshot for G02-Q.1. }
    \label{banner_ex}
\end{figure}

Q.1 While browsing the Web, have you ever come across any banners (see ~\autoref{banner_ex}) that contain information about the use of ``Cookies"?
\begin{itemize}
    \item Yes
    \item No
    \item Not paid attention
\end{itemize}

Q.2 On encountering a cookie banner while browsing the Web, how often do you click ``Accept", ``Consent,'' ``Agree,'' ``Allow,'' ``Confirm,'' or similar meaning buttons on cookie banners? 
\label{G02Q02}

\begin{itemize}
    \item Always
    \item Only sometimes
    \item Mostly not
    \item Never
\end{itemize}

\begin{figure}
    \centering
    \includegraphics[width=0.75\linewidth]{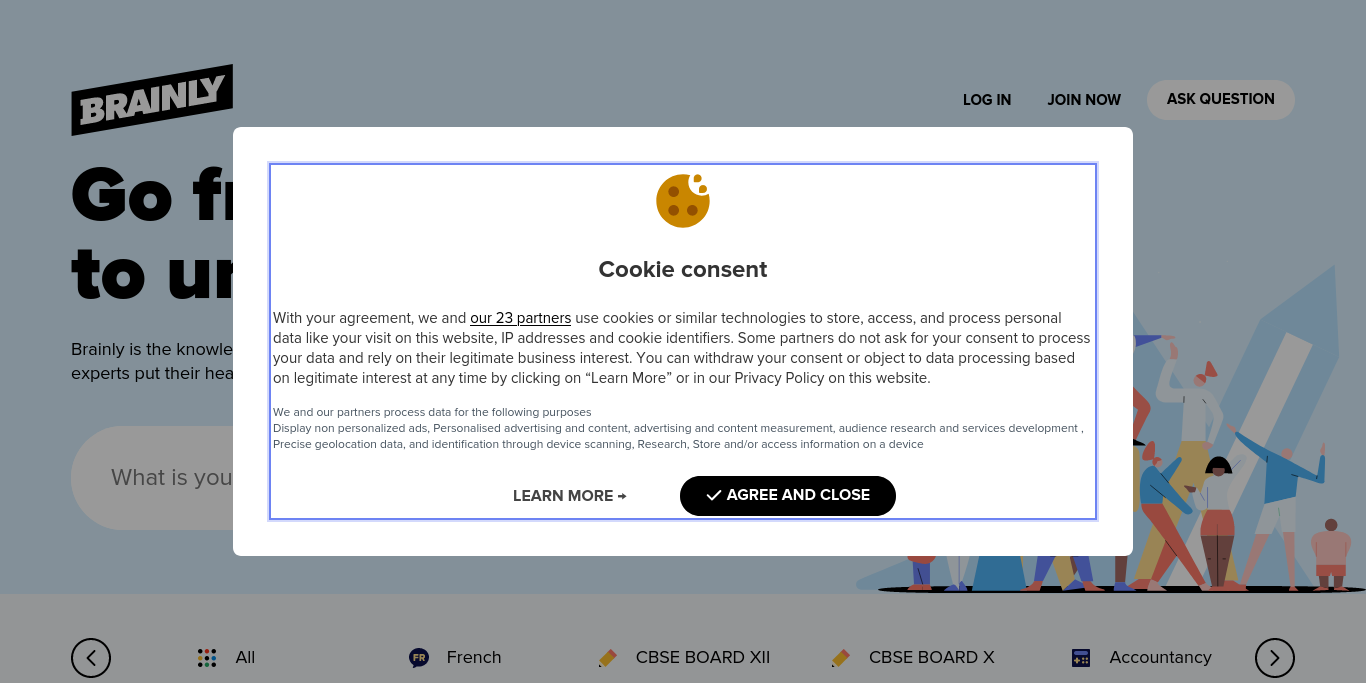}
    \caption{Banner screenshot for G02-Q.3 (B1.1).}
    \label{fig:b1}
\end{figure}

\begin{figure}
    \centering
    \includegraphics[width=0.75\linewidth]{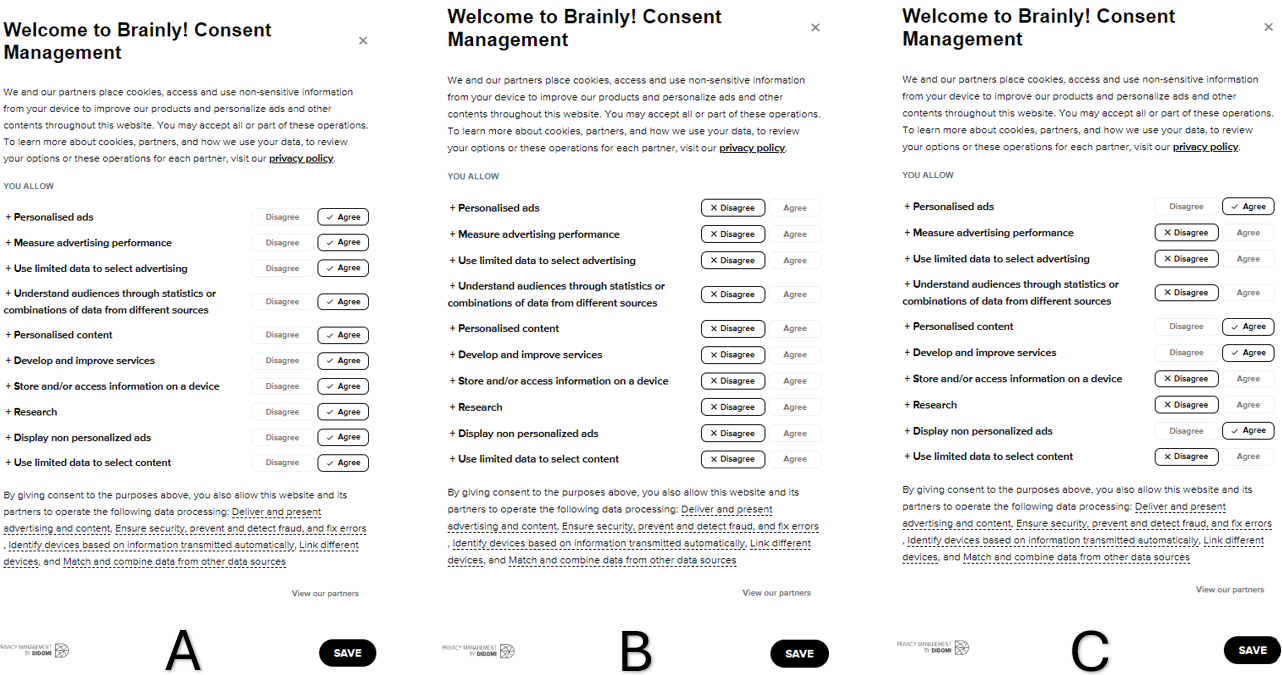}
    \caption{Screenshot for G02-Q.4.}
    \label{fig:custm}
\end{figure}

Q.3 Suppose you encounter this cookie banner (see ~\autoref{fig:b1}); how do you interact with it, given that you cannot browse the website unless you choose one of the options? \label{G02Q03}
\begin{itemize}
    \item I click on ``Agree and Close"
    \item I click on ``Learn More"
    \item I exit the website
\end{itemize}
Q.4 Suppose you clicked on ``Learn More", which of the following settings you go for? (see ~\autoref{fig:custm})
\begin{itemize}
    \item A
    \item B
    \item C
    \item Some other customization
\end{itemize}

\begin{figure}
    \centering
    \includegraphics[width=0.75\linewidth]{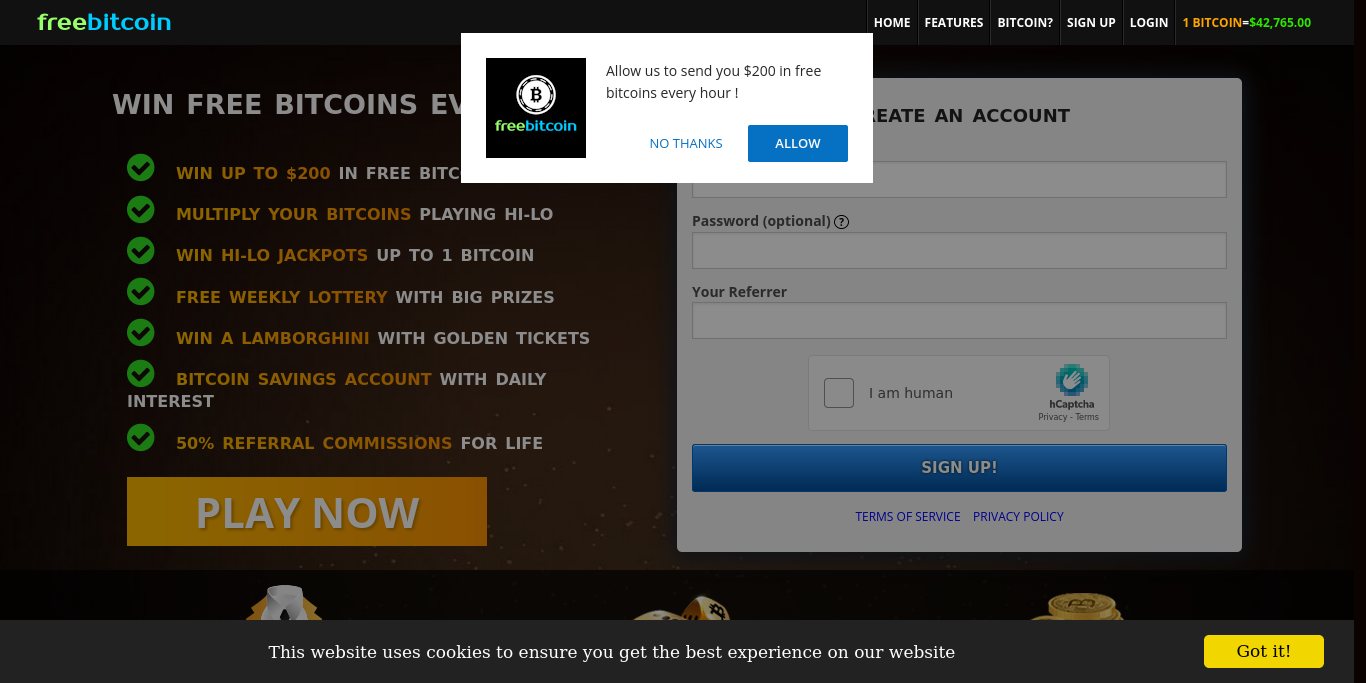}
    \caption{Banner screenshot for G02-Q.5 (B3).}
    \label{fig:b2}
\end{figure}

\begin{figure}
    \centering
    \includegraphics[width=0.75\linewidth]{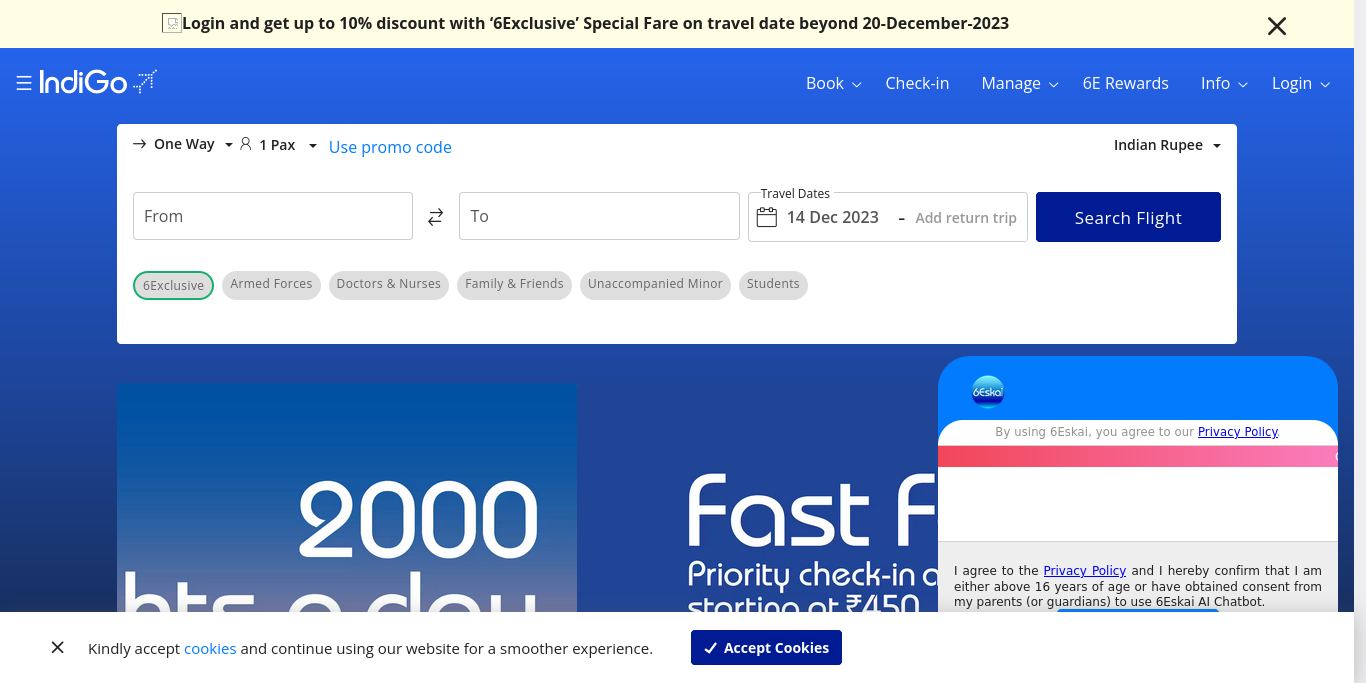}
    \caption{Banner screenshot for G02-Q.6 (B1.2).}
    \label{fig:b3}
\end{figure}

Q.5 Suppose you encounter this cookie banner (see ~\autoref{fig:b2}), how do you interact with it given that you only have one option to select (see bottom of image) and cannot browse the website unless you click on it?
\begin{itemize}
    \item I click in ``Got it!"
    \item I exit the website
    \item I avoid visiting such a website
\end{itemize}
Q.6 Suppose you encounter this cookie banner (see ~\autoref{fig:b3}), how do you interact with it given that it doesn't interrupt your website browsing experience?
\begin{itemize}
    \item I click on ``Accept Cookies'' 
    \item I close the banner 
    \item I ignore the banner and continue browsing 
\end{itemize}

Q.7 On encountering a cookie banner while browsing the web, how often do you click ``Reject,'' ``Decline," ``Disagree,'' ``Refuse,'' or ``Deny'' or similar meaning buttons on cookie banners? \label{G02Q07}
\begin{itemize}
    \item Always
    \item Only sometimes
    \item Mostly not
    \item Never
\end{itemize}

\begin{figure}
    \centering
    \includegraphics[width=0.75\linewidth]{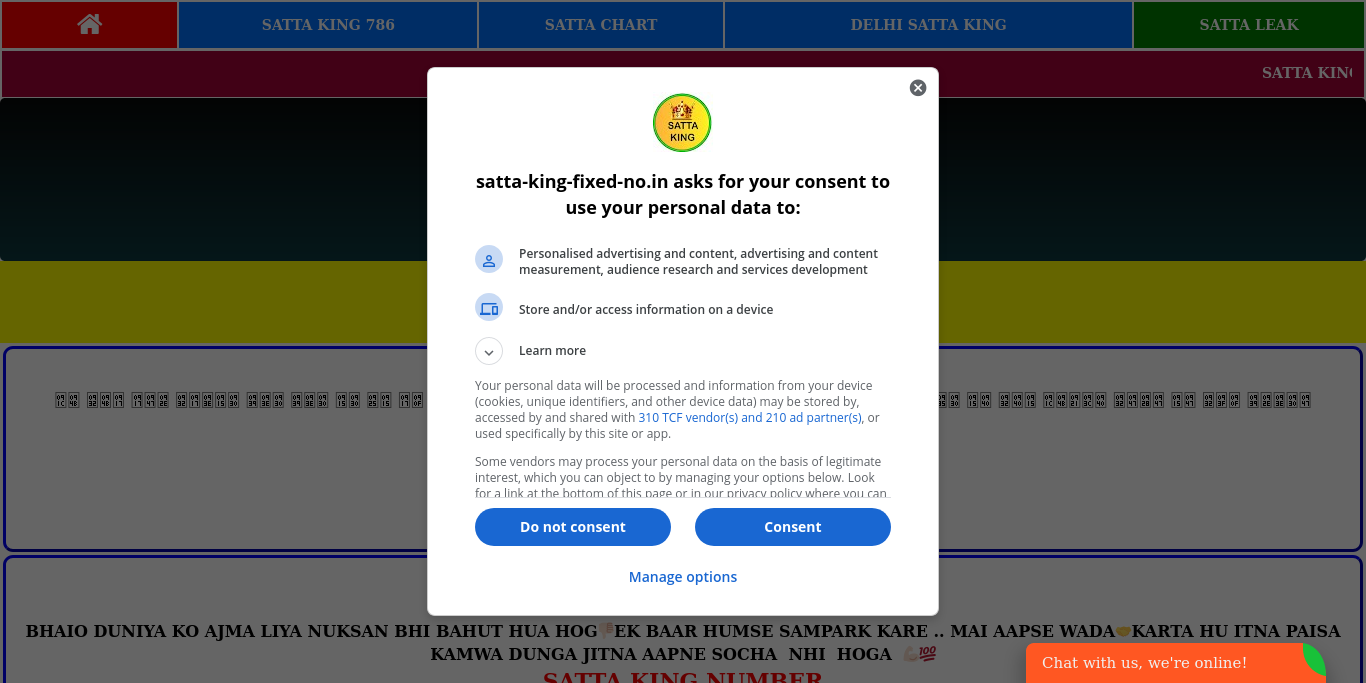}
    \caption{Banner screenshot for G02-Q.8 (B2).}
    \label{fig:b4}
\end{figure}

Q.8 Suppose you encounter this cookie banner (see \autoref{fig:b4}); how do you interact with it, given that you can close it and then continue browsing the website? 
\label{G02Q08}

\begin{itemize}
    \item I click on ``Consent'' 
    \item I click on ``Do not consent'' 
    \item I click on ``Manage options'' 
    \item I close the banner 
\end{itemize}
Q.9 From your experience, how easy is it to find and click an ``Accept'' button on cookie banners?  
\begin{itemize}
    \item Very easy
    \item Easy
    \item Difficult
    \item Very difficult

\end{itemize}
Q.10 From your experience, how easy is it to find and click a ``Reject'' Button on cookie banners?  
\begin{itemize}
    \item Very easy
    \item Easy
    \item Difficult
    \item Very difficult
\end{itemize}

\noindent\hrulefill 

\subsection*{\textbf{Group 3: Your Perspective (G03)}}
\label{group3}
Q.1 Privacy of my browsing history is \textit{very} important to me. \label{G03Q01}
\begin{itemize}
    \item Strongly agree
    \item Agree
    \item Disagree 
    \item Strongly disagree
\end{itemize}
Q.2 My browsing history is \textit{not} safe on the Internet. \label{G03Q02}
\begin{itemize}
    \item Strongly agree
    \item Agree
    \item Disagree
    \item Strongly disagree
\end{itemize}
Q.3 The websites I visit \textit{do not} use tracking or profiling techniques. \label{G03Q03}
\begin{itemize}
    \item Strongly agree
    \item Agree
    \item Disagree
    \item Strongly disagree

\end{itemize}
Q.4 I am fine with websites using \textit{third-party cookies} for tracking/profiling me across the Internet.\\ (\emph{Note: Third-party cookies are data stored by websites other than the one you visited, often used for tracking your activity across multiple sites, enabling targeted advertising.}) \label{G03Q04}
\begin{itemize}
    \item Strongly agree
    \item Agree
    \item Disagree
    \item Strongly disagree
\end{itemize}
Q.5 Cookie banners provide exact and true information about the purpose for \textit{collecting/processing/storing} my data.\label{G03Q05}
\begin{itemize}
    \item Strongly agree
    \item Agree
    \item Disagree
    \item Strongly disagree
\end{itemize}
Q.6 cookie banners respect my \textit{consent/denial} for setting cookies. \label{G03Q06}
\begin{itemize}
    \item Strongly agree
    \item Agree
    \item Disagree
    \item Strongly disagree
\end{itemize}
Q.7 \textit{Right to Personal Data Privacy} is an unnecessary fundamental right in India. \label{G03Q07}
\begin{itemize}
    \item Strongly agree
    \item Agree
    \item Disagree
    \item Strongly disagree
\end{itemize}
Q.8 India needs a standalone comprehensive Personal-Data Privacy law to regulate personal data protection. \label{G03Q08} 
\begin{itemize}
    \item Strongly agree
    \item Agree
    \item Disagree
    \item Strongly disagree
\end{itemize}

\noindent\hrulefill 

\subsection*{\textbf{Group 4: Your Acceptance (G04)}}
\label{group4}
Q.1 Provisions of DPDP Act shall not apply to the State for personal data processing in the interests of sovereignty and integrity of India, security of the State, friendly relations with foreign States, maintenance of public order or preventing incitement to any cognizable offense relating to any of these. As a citizen, I find this provision - \label{G04Q01}            
\newline 
\emph{[Chapter-IV, Section 17, Sub-section 2, Clause (a)]}
\begin{itemize}
    \item Highly acceptable
    \item Acceptable
    \item Unacceptable
    \item Highly unacceptable
\end{itemize}

Q.2 Citizens' consent may not be required for processing their personal data for specified\textit{ legitimate purposes} (as defined by the Central Government) such as voluntary sharing of data by the individual or processing by the State for permits, licenses, benefits, and services. As a citizen, I find this provision - 
 \label{G04Q02}         \newline 
\emph{[Chapter-II, Section 7, Clause (b)]  }
\begin{itemize}
    \item Highly acceptable
    \item Acceptable
    \item Unacceptable
    \item Highly unacceptable
\end{itemize}

Q.3 The Central Government may, by notification, exempt certain activities from the application of DPDP's enforcement, including research, archiving, or statistical purposes. As a citizen, I find this provision -\label{G04Q03}               \newline 
\emph{[Chapter-IV, Section 17, Sub-section 2, Clause (b)]}
\begin{itemize}
    \item Highly acceptable
    \item Acceptable
    \item Unacceptable
    \item Highly unacceptable
\end{itemize}

Q.4 The Central Government may exempt certain data fiduciaries and startups from some obligations of the DPDP act based on the nature and volume of data they process. One of the obligations could be to not present a clear notice to citizens stating the nature of the data collected and the purpose of processing it while asking for their consent. As a citizen, I find this provision - \label{G04Q04}                       \newline 
\emph{[Chapter-IV, Section 17, Sub-section 3]}
\begin{itemize}
    \item Highly acceptable
    \item Acceptable
    \item Unacceptable
    \item Highly unacceptable
\end{itemize}

Q.5 The Data Protection Board of India will be in charge of imposing penalties on businesses/individuals who violate the DPDP law, and the sum of such penalties will be credited to the consolidated fund of India. But there is no compensation stated anywhere for the affected citizens in the law. As a citizen, I find this provision - \newline 
\emph{[Chapter-VIII, Section 34]}
\begin{itemize}
    \item Highly acceptable
    \item Acceptable
    \item Unacceptable
    \item Highly unacceptable
\end{itemize}

Q.6 If a citizen has voluntarily made available their personal data on the Internet, such as blogging their views on a social media platform, then the provisions and protection clauses of DPDP do not apply. As a citizen, I find this provision-\newline 
\emph{[Chapter I, Section 3, Clause (c), Part (ii)]}
\begin{itemize}
    \item Highly acceptable
    \item Acceptable
    \item Unacceptable
    \item Highly unacceptable
\end{itemize}

Q.7 The Chairperson and the Members of the Data Protection Board of India will be appointed by the Central Government alone. As a citizen, I find this provision -\newline 
\emph{[Chapter-V, Section 19, Sub-section 2]}
\begin{itemize}
    \item Highly acceptable
    \item Acceptable
    \item Unacceptable
    \item Highly unacceptable
\end{itemize}

Q.8 All members (including the Chairperson) of the Data Protection Board of India will be appointed for a period of 2 years with scope for re-appointment. As a citizen, I find this provision - \newline 
\emph{[Chapter-V, Section 20, Sub-section 2]}
\begin{itemize}
    \item Highly acceptable
    \item Acceptable
    \item Unacceptable
    \item Highly unacceptable
\end{itemize}

Q.9 No lawsuit, prosecution, or other legal proceedings can be filed against the Central Government, the board, its chairperson, and any member, officer, or employee for anything which is done or intended to be done \textit{in good faith} under the provisions of the DPDP Act.
As a citizen, I find this provision - \newline 
\emph{[Chapter-IX, Section 35]}
\begin{itemize}
    \item Highly acceptable
    \item Acceptable
    \item Unacceptable
    \item Highly unacceptable
\end{itemize}

Q.10 The DPDP Act allows the transfer of personal data outside India for \textit{legitimate purposes} (as defined by the Central Government). \textit{However,} transfer to all countries is not restricted by default but only to the restricted countries notified by the Central Government. As a citizen, I find this provision -\newline 
\emph{[Chapter-IV, Section 16, Sub-section 1]}
\begin{itemize}
    \item Highly acceptable
    \item Acceptable
    \item Unacceptable
    \item Highly unacceptable
\end{itemize}

\noindent\hrulefill 

\subsection*{\textbf{Open-ended question}}
\label{group5}
Please share final thoughts/comments regarding this survey, our research goals, our methodology, or the DPDP Act: \newline
(Open-field) 

\subsection{Codebook:} \label{codebook}
\begin{itemize}
    \item Good and timely survey (17)
    \item Data protection is important (14)
    \item Relevant research (8)
    \item Privacy is important (6)
    \item Distrust in the government (6)
    \item DPDPA needs revision (6) 
    \item Pessimistic about DPDPA (5) 
    \item Government should safeguard user data (5) 
    \item User consent is important (5) 
    \item Cookies are problematic (5)
    \item National security vs individual privacy (4) 
    \item Informative survey (4) 
    \item Chairperson appointment process needs modification (3) 
    \item Need for data protection regulation (3) 
    \item Punishment for non-compliance (3) 
    \item DPDPA not good enough (3) 
    \item No exemption for government (3) 
    \item More stakeholders' involvement (3) 
    \item Data leakage is disastrous (3) 
    \item Privacy is at risk already (3)
    \item Awareness about DPDPA/online privacy (3) 
    \item Awareness oriented survey (3)
    \item Unaware of DPDPA (3) 
    \item Need for compensation for affected citizens (2) 
    \item Unclear benefits for citizens (2) 
    \item User digital literacy (2) 
    \item Need for transparency (2) 
    \item Learn from global good practices (2) 
    \item DPDPA promotes privacy (2) 
    \item Incorporation of rejection options (2) 
    \item DPDPA needs clarification (2)
    \item DPDPA in nascent stage (2)
    \item Implement DPDPA (2)
    \item Good inclusivity of participants (2)
    \item DPDPA should protect users (1)
    \item Supportive of authoritarianism (1) 
    \item Supportive of nationalism (1)
    \item Trust in judiciary (1)
    \item Concerned about privacy (1)
    \item Autonomy (1)
    \item DPDPA addresses existing needs (1)
    \item DPDPA necessitates how businesses approach privacy (1) 
    \item Responsible handling of user data (1)
    \item Government incapable of protecting user data (1)
    \item Trust in government (1)
    \item DPDPA has room for improvement (1)
    \item Partial acceptance of DPDPA (1)
    \item Lack of rejection option (1) 
\end{itemize}

\subsection{Flyer} \label{flyer}
 
\begin{figure}[H]
    \centering
    \includegraphics[width=1\linewidth]
    {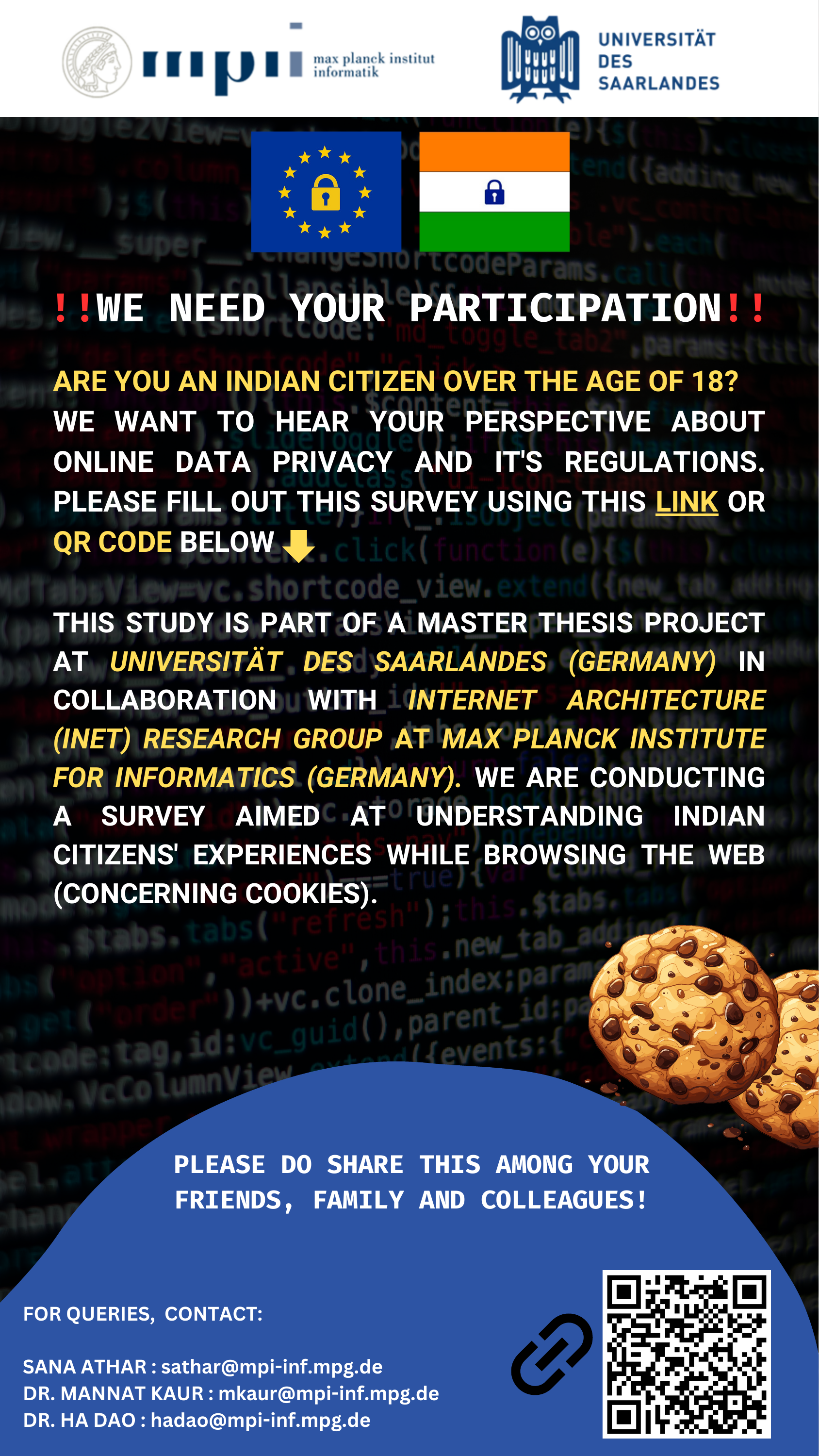}
    \caption{Survey flyer}
    \label{fig:enter-label}
\end{figure}

\end{document}